\documentclass[12pt,a4paper]{article}

\usepackage{amsmath,amsfonts,amssymb,bm}

\usepackage{graphicx} 

\usepackage[hang,normalsize,bf]{subfigure} 

\begin{document}

\begin{center}

{\Huge \bf
Unintegrated CCFM \\ parton distributions \\
and pion production \\ in proton-proton collisions
  \\ at high energies 
}

\vspace {0.6cm}

{\large M. Czech $^{1,2}$ and A. Szczurek $^{1,3}$}

\vspace {0.2cm}

$^{1}$ {\em Institute of Nuclear Physics PAN\\
PL-31-342 Cracow, Poland\\}
$^{2}$ {\em Institute of Physics, Jagiellonian University\\
PL-30-059 Cracow, Poland\\}
$^{3}$ {\em University of Rzesz\'ow\\
PL-35-959 Rzesz\'ow, Poland\\}

\end{center}

\begin{abstract}
Inclusive cross sections for pion production in proton-proton
collisions are calculated for the first time based on
unintegrated parton (gluon, quark, antiquark) distributions (uPDF).
We use Kwieci\'nski uPDF's and
phenomenological fragmentation functions from the literature.
In addition to the $gg \to g$ diagram used recently for applications
at RHIC we include also $g q \to q$ and $q g \to q$ diagrams.
We find that the new contributions are comparable to
the purely gluonic one at midrapidities
and dominate in the fragmentation region.
The new mechanisms are responsible for $\pi^+ - \pi^-$ asymmetry.
We discuss how the asymmetry depends on $x_F$ and $p_t$.
Inclusive distributions in $x_F$ (or rapidity) and transverse
momentum for partons and pions are shown for illustration.
In contrast to standard collinear approach in our approach
the range of applicability can be extended towards much lower
transverse momenta.
\end{abstract}


\section{Introduction}

The distributions of mesons at large transverse momenta
in $p p$ or $p \bar p$ collisions are usually calculated
in the framework of perturbative QCD using collinear factorization
(see e.g. \cite{Owens,Field,EK97,EH02}). While the shape at
transverse momenta larger than 2-4 GeV can be relatively
well explained, the difference between the data and the lowest-order
computation is quantified in terms of a so-called K-factor,
independent quantity for each energy \cite{EH02}. The K-factor is found
to systematically decrease with growing energy \cite{BFLPZ01}.
In order to extend the calculation towards lower values of
meson transverse momenta it was suggested to add an extra
Gaussian distribution in transverse momentum
\cite{Wang2000,Levai_LO,Levai_NLO}
\footnote{A similar procedure was used e.g. for
prompt photon production \cite{WW98}}.
In this approach the standard collinear integrals are replaced as:
\begin{equation}
dx \; p(x,\mu^2) \to dx d^2 k_t \; g(k_t) p(x,\mu^2) \; ,
\label{kt_kick}
\end{equation}
where the extra distribution function in transverse momentum
is normalized to unity
\begin{equation}
\int d^2 k_t \; g(k_t) = 1 .
\label{normalization}
\end{equation}
It is customary to use Gaussian distributions for $g(k_t)$.
It becomes clear that this procedure is effective in
the following sense.
The transverse momentum originates either from the nonperturbative
``really internal'' momentum distributions of partons in nucleons
(of the order of a fraction of GeV) and/or is generate dynamically
as the inital state radiation process (of the order of GeV).
In principle, the second component may depend on the values of
longitudinal momentum fractions $x_1$ and/or $x_2$.
The formalism used by us in the present paper will include both
these effects separately and explicitly.

The recent results from RHIC (see e.g. \cite{RHIC}) have attracted
a renewed interest in better understanding the dynamics of
particle production, not only in nuclear collisions.
Quite different approaches have been used to describe the particle
spectra from the nuclear collisions \cite{PHOBOS}.
The model in Ref.\cite{KL01} with an educated guess
for UGD describes surprisingly well the whole charged particle
rapidity distribution by means of gluonic mechanisms only.
Such a gluonic mechanism would lead to identical production
of positively and negatively charged hadrons.
The recent results of the BRAHMS experiment concerning
heavy ion collisions \cite{BRAHMS} show that the $\pi^-/\pi^+$
and $K^-/K^+$ ratios differ from unity. This put into
question the successful description of Ref.\cite{KL01}.
In the light of this experiment, it becomes obvious that
the large rapidity regions have more complicated flavour structure.
At lower energies these ratios are known to differ from unity
drastically \cite{Antreasyan}.

In Ref.\cite{szczurek03} one of us has calculated inclusive pion
spectra in proton-proton collisions based on different models of
unintegrated gluon distributions taken from the literature.
In the present paper in addition to the $gg \to g$ mechanism
we include also $q_f g \to q_f$ and $g q_f \to q_f$ mechanisms
and similar ones for antiquarks in order to obtain a fully
consistent description.

Many unintegrated gluon distributions in the literature
are ad hoc parametrizations of different sets of
experimental data rather than derived from QCD.
An example of a more systematic approach, making use of familiar
collinear distributions can be found in Ref.\cite{KMR01}.
Recently Kwieci\'nski and collaborators
\cite{CCFM_b1,CCFM_b2,GKB03} have shown how to solve the so-called
CCFM equations by introducing unintegrated parton distributions in
the space conjugated to the transverse momenta \cite{CCFM_b1}.
We present first results for pion production based on unintegrated
parton (gluon, quark, antiquark) distributions obtained by solving
a set of coupled equations developed by Kwieci\'nski
and collaborators.
Recently these parton distributions were tested
for inclusive gauge boson production in proton-antiproton
collisions \cite{KS04} and for charm-anticharm correlations
in photoproduction \cite{LS04}. While in the first process
one tests mainly quark and antiquark distributions at scales
$\mu^2 \sim M_W^2, M_Z^2$, in the second reaction one tests
mainly gluon distributions at scales $\mu^2 \sim m_c^2$. 
In comparison to those reactions
in the present application one tests both gluon as well as
quark and antiquark distributions in a ``more nonperturbative''
region of smaller scales of the order down to
$\mu \sim p_t$(parton) $\sim$ 0.5 - 1.0 GeV,
which corresponds to pion transverse momenta
$p_t$(pion) $\sim$ 0.25 - 0.5 GeV.
This is a region where perturbative and nonperturbative effects
are believed to mix up and the application of the pQCD is doubtful.
On the other hand, this is an interesting region of phase space
responsible for the bulk of hadronic production.
We shall discuss how far down to small pion transverse momenta
one can apply the present approach.

Some preliminary results of the present study were presented at
a conference \cite{preliminary}.

\section{Kwieci\'nski unintegrated parton distributions}

Kwieci\'nski has shown that the evolution equations
for unintegrated parton distributions takes a particularly
simple form in the variable conjugated to the parton transverse momentum.
The two possible representations are interrelated via Fourier-Bessel
transform
\begin{equation}
  \begin{split}
    &{f_k(x,\kappa_t^2,\mu^2)} =
    \int_{0}^{\infty} db \;  b J_0(\kappa_t b)
    {{\tilde f}_k(x,b,\mu^2)} \; ,
    \\
    &{{\tilde f}_k(x,b,\mu^2)} =
    \int_{0}^{\infty} d \kappa_t \;  \kappa_t J_0(\kappa_t b)
    {f_k(x,\kappa_t^2,\mu^2)} \; .
  \end{split}
\label{Fourier}
\end{equation}
The index k above numerates either gluons (k=0), quarks (k$>$ 0) or
antiquarks (k$<$ 0).
In the impact-parameter space the Kwieci\'nski equation
takes the following simple form
\begin{equation}
\begin{split}
{\partial{\tilde f_{NS}(x,b,\mu^2)}\over \partial \mu^2} &=
{\alpha_s(\mu^2)\over 2\pi \mu^2}  \int_0^1dz  \, P_{qq}(z)
\bigg[\Theta(z-x)\,J_0((1-z) \mu b)\,
{\tilde f_{NS}\left({x\over z},b,\mu^2 \right)}
\\&- {\tilde f_{NS}(x,b,\mu^2)} \bigg]  \; , \\
{\partial{\tilde f_{S}(x,b,\mu^2)}\over \partial \mu^2} &=
{\alpha_s(\mu^2)\over 2\pi \mu^2} \int_0^1 dz
\bigg\{\Theta(z-x)\,J_0((1-z) \mu b)\bigg[P_{qq}(z)\,
 {\tilde f_{S}\left({x\over z},b,\mu^2 \right)}
\\&+ P_{qg}(z)\, {\tilde f_{G}\left({x\over z},b,\mu^2 \right)}\bigg]
 - [zP_{qq}(z)+zP_{gq}(z)]\,
{\tilde f_{S}(x,b,\mu^2)}\bigg\}  \; ,
 \\
{ \partial {\tilde f_{G}(x,b,\mu^2)}\over \partial \mu^2}&=
{\alpha_s(\mu^2)\over 2\pi \mu^2} \int_0^1 dz
\bigg\{\Theta(z-x)\,J_0((1-z) \mu b)\bigg[P_{gq}(z)\,
{\tilde f_{S}\left({x\over z},b,\mu^2 \right)}
\\&+ P_{gg}(z)\, {\tilde f_{G}\left({x\over z},b,\mu^2 \right)}\bigg]
-[zP_{gg}(z)+zP_{qg}(z)]\, {\tilde f_{G}(x,b,\mu^2)}\bigg\} \; .
\end{split}
\label{kwiecinski_equations}
\end{equation}
We have introduced here the short-hand notation
\begin{equation}
\begin{split}
\tilde f_{NS}&= \tilde f_u - \tilde f_{\bar u}, \;\;
                 \tilde f_d - \tilde f_{\bar d} \; ,  \\
\tilde f_{S}&= \tilde f_u + \tilde f_{\bar u} + 
                \tilde f_d + \tilde f_{\bar d} + 
                \tilde f_s + \tilde f_{\bar s} \; . 
\end{split}
\label{singlet_nonsinglet}
\end{equation}
The unintegrated parton distributions in the impact factor
representation are related to the familiar collinear distributions
as follows
\begin{equation}
\tilde f_{k}(x,b=0,\mu^2)=\frac{x}{2} p_k(x,\mu^2) \; .
\label{uPDF_coll_1}
\end{equation}
On the other hand, the transverse momentum uPDF's are related to the
integrated distributions as
\begin{equation}
x p_k(x,\mu^2) =
\int_0^{\infty} d \kappa_t^2 \; f_k(x,\kappa_t^2,\mu^2) \; .
\label{uPDF_coll_2}
\end{equation}
While physically $f_k(x,\kappa_t^2,\mu^2)$ should be positive,
there is no obvious reason for such a limitation for
$\tilde f_k(x,b,\mu^2)$.

In the following we use leading-order parton distributions
from ref.\cite{GRV98} as the initial condition for QCD evolution.
The set of integro-differential equations in b-space
was solved by the method based on the discretisation made with
the help of the Chebyshev polynomials (see e.g. \cite{GKB03}).
Then the unintegrated parton distributions were put on a grid
in $x$, $b$ and $\mu^2$ and the grid was used in practical
applications for Chebyshev interpolation (see next section). 

\section{Inclusive cross sections for partons}

The approach proposed by Kwieci\'nski is very convenient to
introduce the nonperturbative effects like
internal (nonperturbative) transverse momentum distributions
of partons in nucleons.
It seems reasonable, at least in the first approximation,
to include the nonperturbative effects in the factorizable way
\begin{equation}
\tilde{f}_q(x,b,\mu^2) = 
\tilde{f}_q^{CCFM}(x,b,\mu^2)
 \cdot F_q^{np}(b) \; .
\label{modified_uPDFs}
\end{equation}
The form factor responsible for the nonperturbative effects
must be normalized such that
\begin{equation}
F^{NP}(b=0) = 1
\label{ff_normalization}
\end{equation}
in order not to spoil the relation (\ref{uPDF_coll_1}).
In the following, for simplicity, we use a flavour and
$x$-independent form factor
\begin{equation}
F_q^{np}(b) = F^{np}(b) = \exp\left(-\frac{b^2}{4 b_0^2}\right) \; 
\label{formfactor}
\end{equation}
which describes the nonperturbative effects.
The Gaussian form factor in $b$ means also a Gaussian initial
momentum distribution $\propto \exp(-k_t^2 b_0^2)$ (Fourier transform of
a Gaussian function is a Gaussian function). Gaussian form factor
is often used to correct collinear pQCD calculations for the
so-called internal momenta.
Other functional forms in $b$ are also possible.

\begin{figure}[!thb] 
\begin{center}
\includegraphics[width=5.0cm]{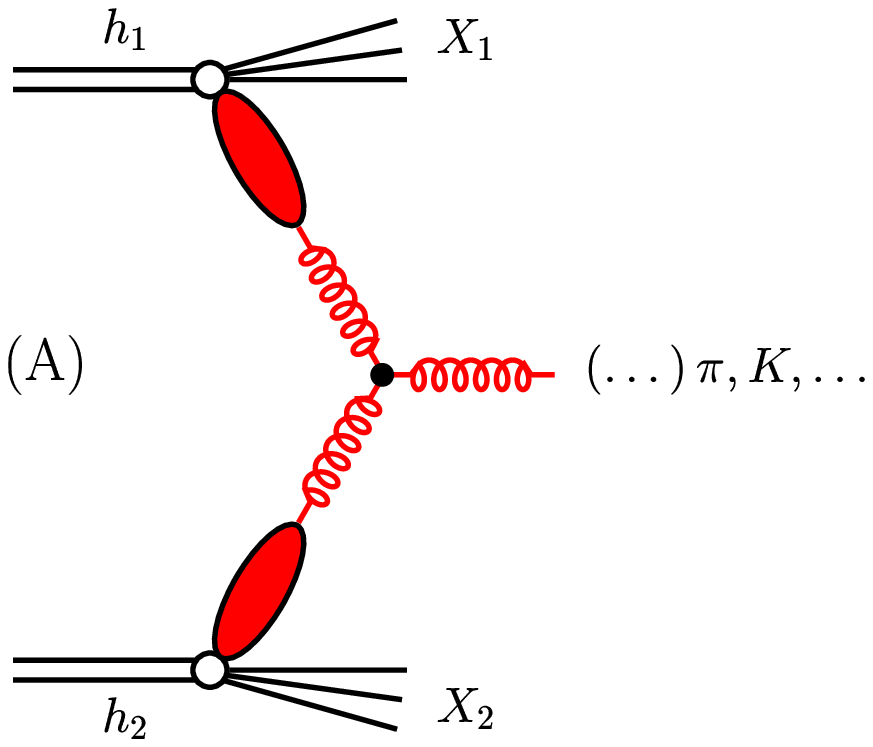}\\[30pt]
\includegraphics[width=5.0cm]{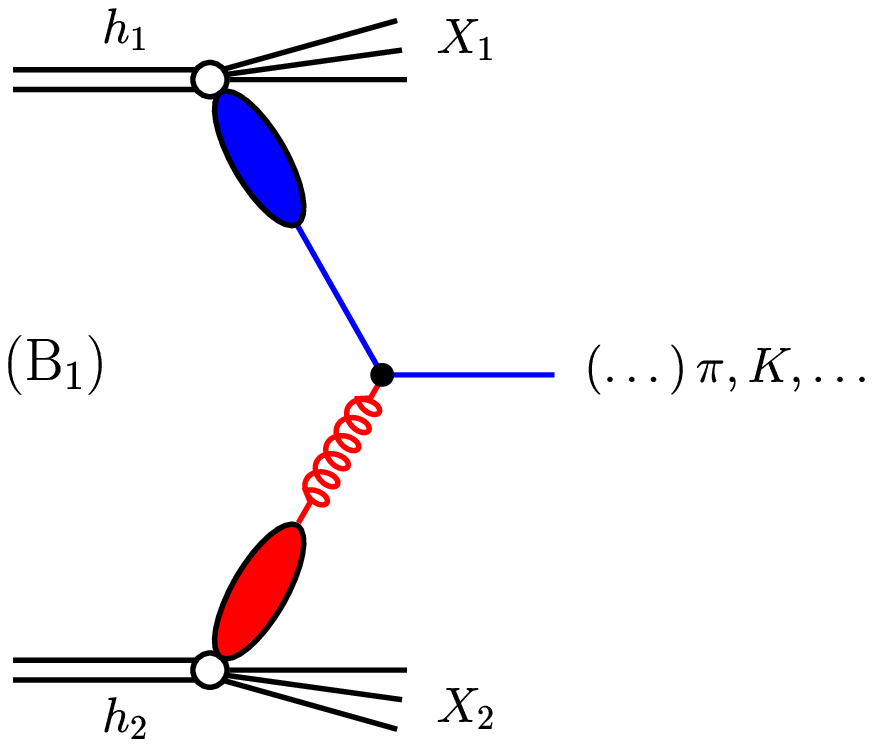}
\includegraphics[width=5.0cm]{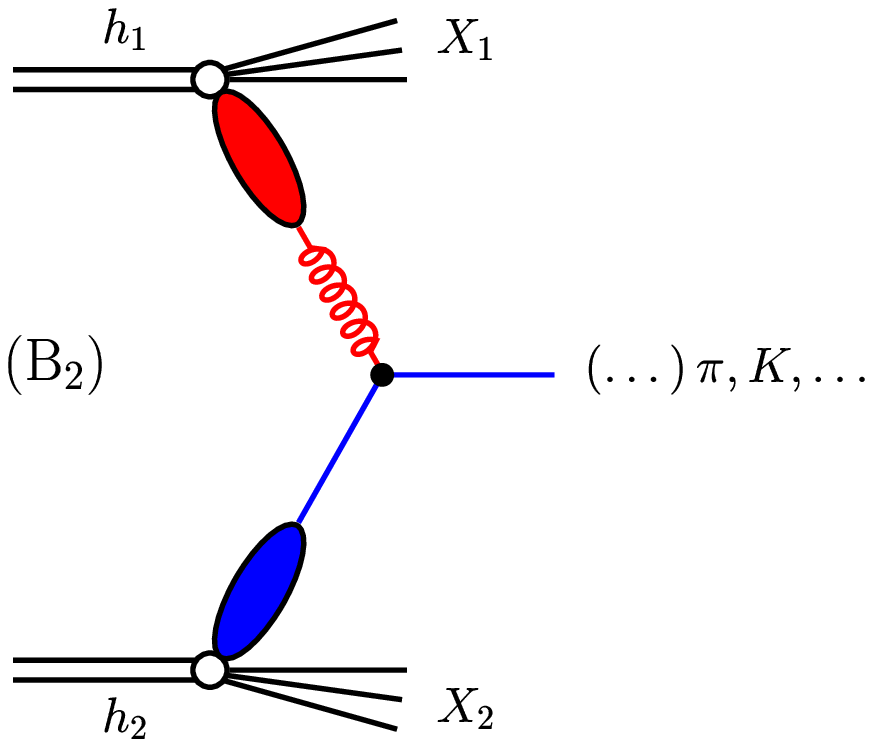}
\caption[*]{Leading-order diagrams for inclusive parton production
\label{fig:diagrams}
}
\end{center}
\end{figure}

The $gg \to g$ mechanism considered in the literature is not
the only one possible. In Fig.\ref{fig:diagrams} we show two other leading
order diagrams. They are potentially important in the so-called
fragmentation region. The formulae for inclusive quark/antiquark
distributions are similar to the formula for $gg \to g$ \cite{GLR81}.
The formulae for all the processes mentioned are listed below:

for diagram A (gg$\to$g):
\begin{eqnarray}
&&\frac{d \sigma^{A}}{dy d^2 p_t} = \frac{16  N_c}{N_c^2 - 1}
{\frac{1}{p_t^2}}
 \nonumber \\
&& \int
 \alpha_s({\Omega^2}) \;
{  f_{g/1}(x_1,\kappa_1^2,\mu^2)} \;
{  f_{g/2}(x_2,\kappa_2^2,\mu^2)}
\nonumber \\
&&\delta^{(2)}(\vec{\kappa}_1+\vec{\kappa}_2 - \vec{p}_t)
\; d^2 \kappa_1 d^2 \kappa_2    \; ,
\label{diagram_A_tr}
\end{eqnarray}

for diagram B$_1$ (q$_f$ g $\to$ q$_f$):
\begin{eqnarray}
&&\frac{d \sigma^{B_1}}{dy d^2 p_t} = \frac{16  N_c}{N_c^2 - 1}
\left( \frac{4}{9} \right)
{\frac{1}{p_t^2}}
 \nonumber \\
&& \sum_f \int
 \alpha_s({\Omega^2}) \;
{  f_{q_f/1}(x_1,\kappa_1^2,\mu^2)} \;
{  f_{g/2}(x_2,\kappa_2^2,\mu^2)}
\nonumber \\
&&\delta^{(2)}(\vec{\kappa}_1+\vec{\kappa}_2 - \vec{p}_t)
\; d^2 \kappa_1 d^2 \kappa_2    \; ,
\label{diagram_B1_tr}
\end{eqnarray}

for diagram B$_2$ (g q$_f$ $\to$ q$_f$):
\begin{eqnarray}
&&\frac{d \sigma^{B_2}}{dy d^2 p_t} = \frac{16  N_c}{N_c^2 - 1}
\left( \frac{4}{9} \right)
{\frac{1}{p_t^2}}
 \nonumber \\
&& \sum_f \int
 \alpha_s({\Omega^2}) \;
{  f_{g/1}(x_1,\kappa_1^2,\mu^2)} \;
{  f_{q_f/2}(x_2,\kappa_2^2,\mu^2)}
\nonumber \\
&&\delta^{(2)}(\vec{\kappa}_1+\vec{\kappa}_2 - \vec{p}_t)
\; d^2 \kappa_1 d^2 \kappa_2    \; .
\label{diagram_B2_tr}
\end{eqnarray}
These seemingly 4-dimensional integrals can be written
as 2-dimensional integrals after a siutable change
of variables \cite{szczurek03}
\begin{equation}
\int \; ...\; \delta^{(2)}(\vec{\kappa}_1+\vec{\kappa}_2 - \vec{p}_t)
\; d^2 \kappa_1 d^2 \kappa_2 =
\int \; ...\; \frac{d^2 q_t}{4} \; .
\end{equation}
The integrands of these ``reduced'' 2-dimensional integrals in 
$\vec{q}_t = \vec{\kappa_1} - \vec{\kappa_2}$ are
generally smooth functions of $q_t$ and corresponding azimuthal
angle $\phi_{q_t}$.
In the following we use two different prescriptions for
the factorization scale $\mu^2$:
\begin{itemize}
\item $\mu^2 = p_t^2$ with freezing for $p_t^2 < \mu^2$,
\item $\mu^2 = p_t^2 + \mu_0^2$.
\end{itemize}
In Eqs.(\ref{diagram_A_tr}), (\ref{diagram_B1_tr}) and
 (\ref{diagram_B2_tr}) the longitudinal momentum
fractions
\begin{equation}
x_{1/2} = \frac{\sqrt{p_t^2 + m_x^2}}{\sqrt{s}} \exp(\pm y) \; ,
\label{x1_x2}
\end{equation}
where $m_x$ is the effective mass of the parton.

The sums in (\ref{diagram_B1_tr}) and (\ref{diagram_B2_tr})
run over both quarks and antiquarks.
The argument of the running coupling constant $\Omega^2$ above
was not specified explicitly yet.
In principle, it can be $p_t^2$ or a combination of $p_t^2$,
$\kappa_1^2$ and $\kappa_2^2$. In the standard transverse
momentum representation it is reasonable to assume
$\Omega^2 = \min(p_t^2,\kappa_1^2,\kappa_2^2)$
(see e.g. \cite{szczurek03}). In the region of very small $p_t$ 
usually $p_t^2 < \kappa_1^2, \kappa_2^2$ and $\Omega_2 = p_t^2$
is a good approximation.

Assuming for simplicity that $\Omega^2 = \Omega^2(p_t^2)$ or $p_t^2$
(function of transverse momentum squared of the ``produced'' parton,
or simply transverse momentum squared)
and taking the following representation of the $\delta$ function
\begin{equation}
\delta^{(2)}(\vec{\kappa_1}+\vec{\kappa_2}-\vec{p}_t) =
\frac{1}{(2 \pi)^2} \int d^2 b \;
\exp \left[   
(\vec{\kappa_1}+\vec{\kappa_2}-\vec{p}_t) \vec{b}
\right] \; ,
\label{delta_representation}
\end{equation}
the  formulae (\ref{diagram_A_tr}), (\ref{diagram_B1_tr})
and (\ref{diagram_B2_tr}) can be written in the equivalent way
in terms of parton distributions in the space conjugated
to the transverse momentum.
The corresponding formulae read:

for diagram A:
\begin{eqnarray}
&&\frac{d \sigma^{A}}{dy d^2 p_t} = \frac{16  N_c}{N_c^2 - 1}
{\frac{1}{p_t^2}}
 \alpha_s({p_t^2}) \nonumber \\
&& \int
{\tilde f}_{g/1}(x_1,b,\mu^2) \;
{\tilde f}_{g/2}(x_2,b,\mu^2)
 J_0(p_t b) \; 2 \pi b db    \; , 
\label{diagram_A_b}
\end{eqnarray}
for diagram B$_1$:
\begin{eqnarray}
&&\frac{d \sigma^{B_1}}{dy d^2 p_t} = \frac{16  N_c}{N_c^2 - 1}
\left( \frac{4}{9} \right)
{\frac{1}{p_t^2}}
 \alpha_s({p_t^2})
 \nonumber \\
&& \sum_f \int
{\tilde f}_{q_f/1}(x_1,b,\mu^2) \;
{\tilde f}_{g/2}(x_2,b,\mu^2)
 J_0(p_t b) \; 2 \pi b db    \; , 
\label{diagram_B1_b}
\end{eqnarray}
for diagram B$_2$:
\begin{eqnarray}
&&\frac{d \sigma^{B_2}}{dy d^2 p_t} = \frac{16  N_c}{N_c^2 - 1}
\left( \frac{4}{9} \right)
{\frac{1}{p_t^2}}
 \alpha_s({p_t^2})
 \nonumber \\
&& \sum_f \int
{\tilde f}_{g/1}(x_1,b,\mu^2) \;
{\tilde f}_{q_f/2}(x_2,b,\mu^2)
 J_0(p_t b) \; 2 \pi b db    \; . 
\label{diagram_B2_b}
\end{eqnarray}
These are 1-dimensional integrals. The price one has to pay is
that now the integrands are strongly oscillating functions of
the impact factor, especially for large $p_t$.
The formulae (\ref{diagram_A_b}), (\ref{diagram_B1_b})
and (\ref{diagram_B2_b}) are very convenient to directly use
the solutions of the Kwieci\'nski equations discussed
in the previous section.

When extending running $\alpha_s$ to the region of small scales
we use a parameter free analytic model from ref.\cite{SS97}.

\section{From partons to hadrons}

In Ref.\cite{KL01} it was assumed, based on the concept
of local parton-hadron duality, that the rapidity distribution
of particles is identical to the rapidity distribution of gluons.
In the present approach we follow a different approach
which makes use of phenomenological fragmentation functions (FF's).
In the following we assume $\theta_h = \theta_g$.
This is equivalent to $\eta_h = \eta_g = y_g$, where $\eta_h$ and
$\eta_g$ are hadron and gluon pseudorapitity, respectively. Then
\begin{equation}
y_g = \mathrm{arsinh} \left( \frac{m_{t,h}}{p_{t,h}} \sinh y_h \right)
\; ,
\label{yg_yh}
\end{equation}
where the transverse mass $m_{t,h} = \sqrt{m_h^2 + p_{t,h}^2}$.
In order to introduce phenomenological FF's
one has to define a new kinematical variable.
In accord with $e^+e^-$ and $e p$ collisions we define a 
quantity $z$ by the equation $E_h = z E_g$.
This leads to the relation
\begin{equation}
p_{t,g} = \frac{p_{t,h}}{z} J(m_{t,h},y_h) \; ,
\label{ptg_pth}
\end{equation}
where the jacobian $J(m_{t,h},y_h)$ reads 
\begin{equation}
J(m_{t,h},y_h) =
\left( 1 - \frac{m_h^2}{m_{t,h}^2 \cosh^2 y_h} \right)^{-1/2} \; .
\label{J}
\end{equation}
Now we can write a given-type parton contribution to
the single particle distribution
in terms of a parton (gluon, quark, antiquark) distribution
as follows
\begin{eqnarray}
\frac{d \sigma^{p} (\eta_h, p_{t,h})}{d \eta_h d^2 p_{t,h}} =
\int d y_p d^2 p_{t,p} \int 
dz \; D_{p \rightarrow h}(z,\mu_D^2)  \nonumber \\
\delta(y_p - \eta_h) \; 
\delta^2\left(\vec{p}_{t,h} - \frac{z \vec{p}_{t,p}}{J}\right)
\cdot \frac{d \sigma (y_p, p_{t,p})}{d y_p d^2 p_{t,p}} \; .
\label{from_gluons_to_particles}
\end{eqnarray}

Please note that this is not an invariant cross section.
The invariant cross section can be obtained via suitable
variable transformation
\begin{equation}
\frac{d \sigma^{p} (y_h, p_{t,h})}{d y_h d^2 p_{t,h}} =
\left( \frac{\partial(y_h,p_{t,h})}
            {\partial(\eta_h,p_{t,h})} \right)^{-1} \;
\frac{d \sigma^{p} (y_h, p_{t,h})}{d \eta_h d^2 p_{t,h}} \; ,
\label{invariant_cross_section}
\end{equation}
where
\begin{equation}
y_h = \frac{1}{2} \log
 \left[
\frac{\sqrt{\frac{m_h^2+p_{t,h}^2}{p_{t,h}^2} + \sinh^2\eta_h } + \sinh\eta_h }
     {\sqrt{\frac{m_h^2+p_{t,h}^2}{p_{t,h}^2} + \sinh^2\eta_h } - \sinh\eta_h }
 \right] \; .
\label{yh_etay}
\end{equation}

Making use of the $\delta$ function in
(\ref{from_gluons_to_particles})
the inclusive distributions of hadrons (pions, kaons, etc.)
are obtained through a convolution of inclusive distributions
of partons and flavour-dependent fragmentation functions
\begin{eqnarray}
{
\frac{d \sigma(\eta_h,p_{t,h})}{d \eta_h d^2 p_{t,h}} } =
\int_{z_{min}}^{z_{max}} dz \frac{J^2}{z^2}  \nonumber \\
{D_{g \rightarrow h}(z, \mu_D^2)}
{\frac{d \sigma_{g g \to g}^{A}(y_g,p_{t,g})}{d y_g d^2 p_{t,g}}}
 \Bigg\vert_{y_g = \eta_h \atop p_{t,g} = J p_{t,h}/z}
 \nonumber \\
+ \sum_{f=-3}^3{D_{q_f \rightarrow h}(z, \mu_D^2)}
{\frac{d \sigma_{q_f g \to q_f}^{B_1}(y_{q_f},p_{t,q_f})}
{d y_{q_f} d^2 p_{t,q}}}
 \Bigg\vert_{y_q = \eta_h \atop p_{t,q} = J p_{t,h}/z}
 \nonumber \\
+ \sum_{f=-3}^3{D_{q_f \rightarrow h}(z, \mu_D^2)}
{\frac{d \sigma_{g q_f \to q_f}^{B_2}(y_{q_f},p_{t,q_f})}
{d y_{q_f} d^2 p_{t,q}}}
 \Bigg\vert_{y_q = \eta_h \atop p_{t,q} = J p_{t,h}/z}
 \; . 
\label{all_diagrams}
\end{eqnarray}
One dimensional distributions of hadrons can be obtained
through the integration over the other variable.
For example the pseudorapidity distribution is
\begin{equation}
\frac{d \sigma(\eta_h)}{d \eta_h} =
\int d^2 p_{t,h} \;
\frac{d \sigma(\eta_h,p_{t,h})}{d \eta_h d^2 p_{t,h}} \; .
\label{eta_had_distribution}
\end{equation}
There are a few sets of fragmentation functions available in
the literature (see e.g. \cite{BKK95}, \cite{Kretzer2000}).

\section{Results}

As an illustration of the formalism, in the present paper,
we shall show results for energies adequate for CERN SPS
i.e. for the energies at which the missing mechanisms should play
an important role.

Before we adress the distributions of pions we wish to
discuss the inclusive spectra of ``produced'' partons.

\subsection{Parton distributions}

\begin{figure}[htb] 
\begin{center}
    \includegraphics[width=12.0cm]{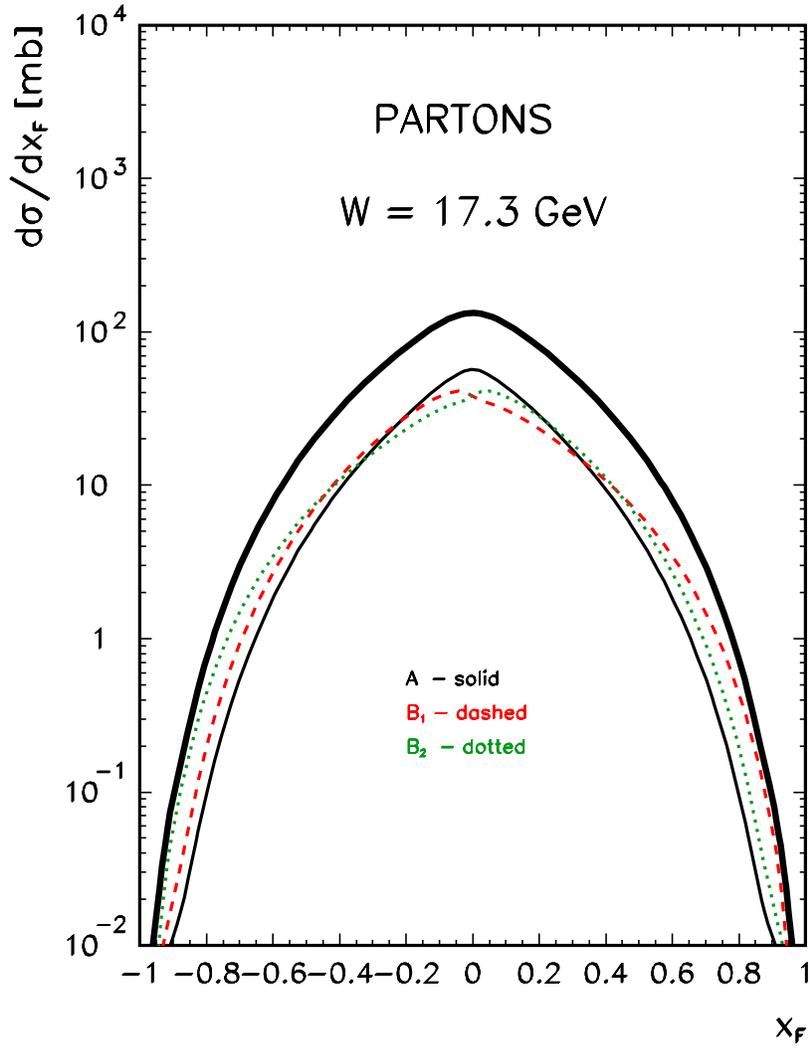}
\caption{\it
Distribution of virtually ``produced'' partons as a function
of $x_F$ for W = 17.3 GeV and $b_0$ = 1 GeV$^{-1}$
of the Gaussian form factor.
In this calculation 0.2 GeV $< p_t <$ 4 GeV.
Contribution of diagram $A$ is shown by thin solid line,
the contribution of diagram $B_1$ by dashed line
and contribution of diagram $B_2$ by dotted line,
and the sum of all processes by thick solid line.
\label{fig:partons_xf}
}
\end{center}
\end{figure}

In the familiar collinear approach the contributions of diagrams
involving quarks and antiquarks are not negligible even at large energies.
A nice, quantitative discussion of this issue can be found
in \cite{EK97}. In this section we shall make a similar
analysis in our approach based on CCFM uPDF's.
In Fig.\ref{fig:partons_xf} we display the contributions of
partons (gluons, quarks, and antiquarks) for all diagrams of
Fig.\ref{fig:diagrams} for the center-of-mass energy W = 17.3 GeV,
i.e. energy of recent experiments of the NA49 collaboration at CERN
as a function of jet (minijet) $x_F$.
The parton rapidity is related to the parton $x_F$ as follows
\begin{equation}
y = \frac{1}{2} \log \left(
\frac{\sqrt{m_x^2 + x_{F}^2 \frac{s}{4} + p_{t}^2} + x_{F} \frac{\sqrt{s}}{2}}
     {\sqrt{m_x^2 + x_{F}^2 \frac{s}{4} + p_{t}^2} - x_{F}
       \frac{\sqrt{s}}{2}}   \; , 
\right) \;
\label{y_x_F}
\end{equation}
where the parton mass here is the same as in Eq.(\ref{x1_x2}).
The corresponding cross section is obtained by integration
over parton transverse momenta in the interval 0.2 GeV $< p_t <$ 4 GeV.
The $gg \to g$ contribution, claimed to be the dominant contribution
at RHIC \cite{KL01}, is somewhat smaller than the contribution
of diagrams $B_1$ and $B_2$. The contribution of diagram $B_1$
(dashed line) dominates at negative Feynman-$x_F$, while the
contribution of diagram $B_2$ (dotted line) at positive Feynman-$x_F$.
By symmetry requirements
$d \sigma^{B_2} / dx_F (x_F) = d \sigma^{B_1} / dx_F(-x_F)$.

\begin{figure}[htb] 
\begin{center}
    \includegraphics[width=12.0cm]{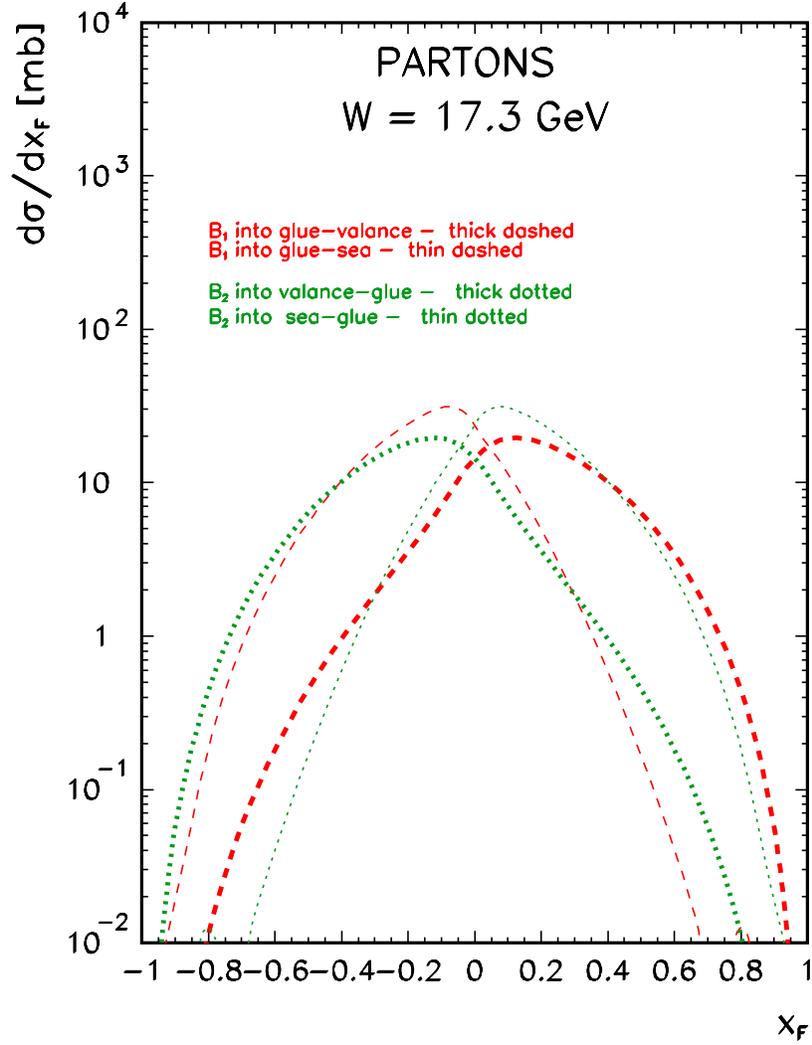}
\caption{\it
Decomposition of diagram $B_1$ cross section into glue-sea
and glue-valence components
and of diagram $B_2$ cross section into sea-glue and valence-glue
components as a function of $x_F$.
\label{fig:B-decomposition_xf}
}
\end{center}
\end{figure}

In order to understand the intriguing asymmetry in $x_F$ of
contributions of diagrams $B_1$ and $B_2$, in
Fig.\ref{fig:B-decomposition_xf} we present a further decomposition
of the $B_1$ contribution into sea-glue, valence-glue subcontributions
and correspondingly of the $B_2$ contribution into glue-sea, glue-valence
subcontributions. This decomposition shows that the sea-glue
and glue-sea contributions are of similar size as the valence-glue
and glue-valence, respectively. It is interesting to note
a different $x_F$-asymmetry of the contributions involving sea
and valence quarks. This decomposition explains the shift of maxima
of contributions corresponding to diagram $B_1$ and $B_2$
seen in Fig.\ref{fig:partons_xf}.

\begin{figure}[htb] 
  \subfigure[]{\label{partons_a_pt}
    \includegraphics[width=7.0cm]{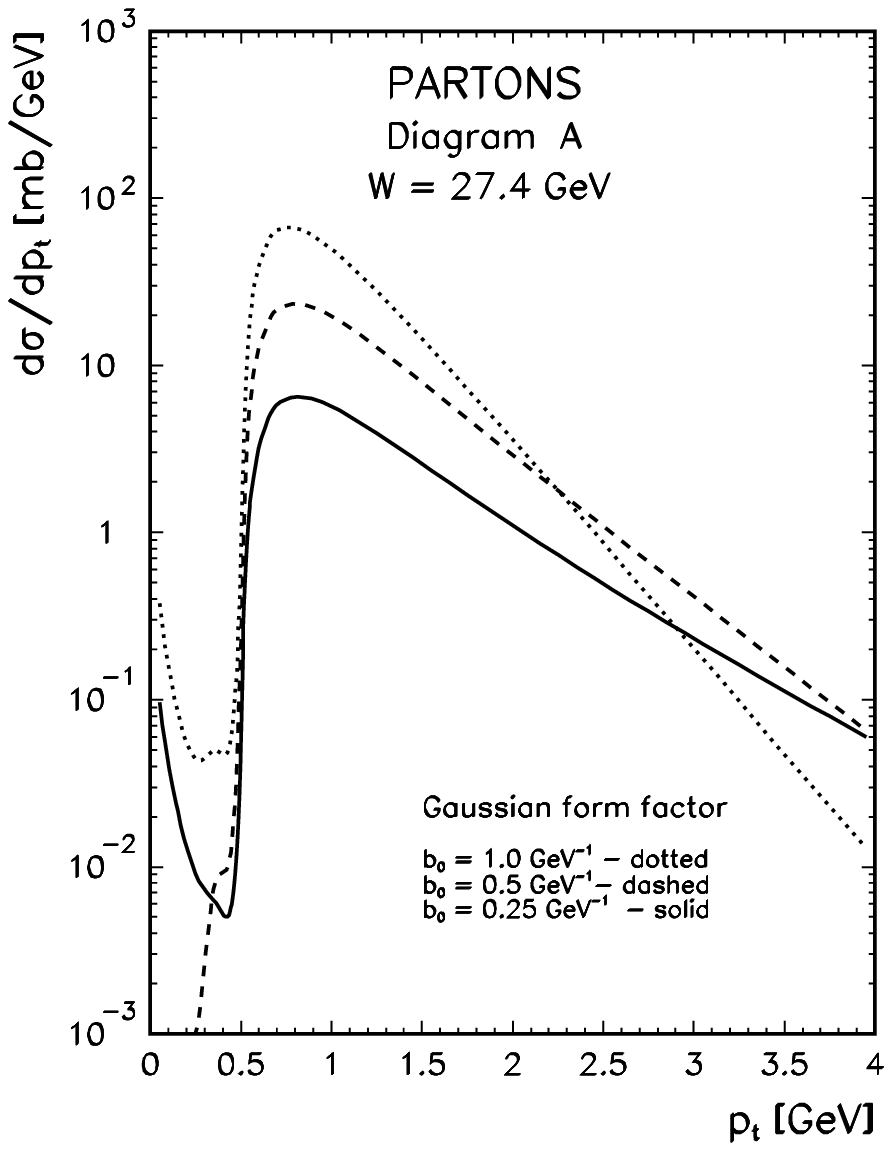}}
  \subfigure[]{\label{partons_b_pt}
    \includegraphics[width=7.0cm]{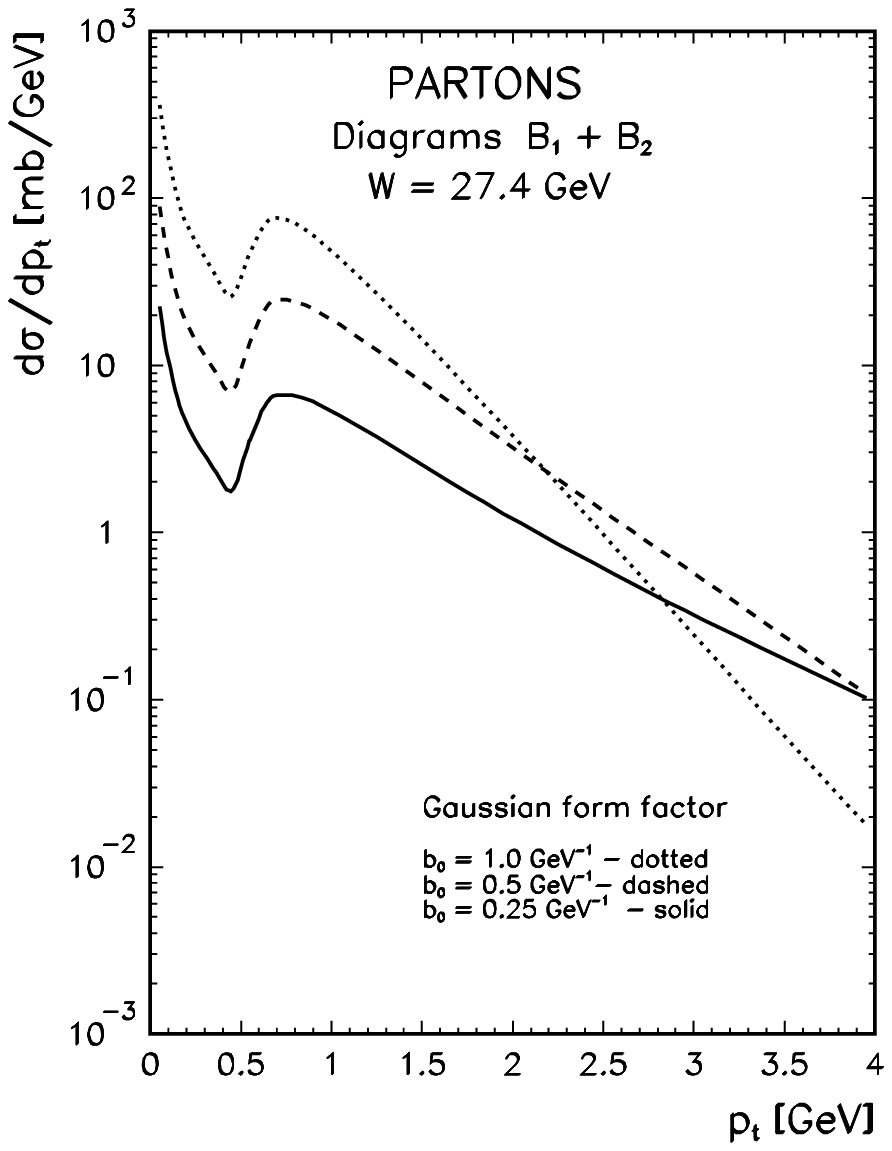}}
\caption{\it
Transverse momentum distribution of partons from diagram $A$ (left panel),
and $B_1 + B_2$ (right panel) for W = 27.4 GeV and different values
of the Gaussian form factor. The cross sections presented were
obtained through integration over -1 $< x_F <$ 1.
\label{fig:partons_pt}
}
\end{figure}

For completeness in Fig.\ref{fig:partons_pt} we show transverse
momentum distribution of ``produced'' gluons corresponding
to diagram $A$ and of quarks and antiquarks corresponding to
diagrams $B_1$ and $B_2$. In this calculation we have integrated over
parton $x_F$ (or parton rapidity).
For the two contributions one obtains
a rather similar functional behaviour.
It is worth noticing that in contrast to the standard collinear
case in our approach the partonic cross section is fully integrable.
This seems promissing in extending the region of applicability
of the ``perturbative''
\footnote{Our approach is not completely perturbative.
Some nonperturbative effects are contained in the nonperturbative
form factor, way of freezing $\alpha_s$ or the scale of parton
distributions.} QCD towards smaller transverse momenta of hadrons.

\begin{figure}[htb] 
  \subfigure[]{\label{partons_a_pt}
    \includegraphics[width=7.0cm]{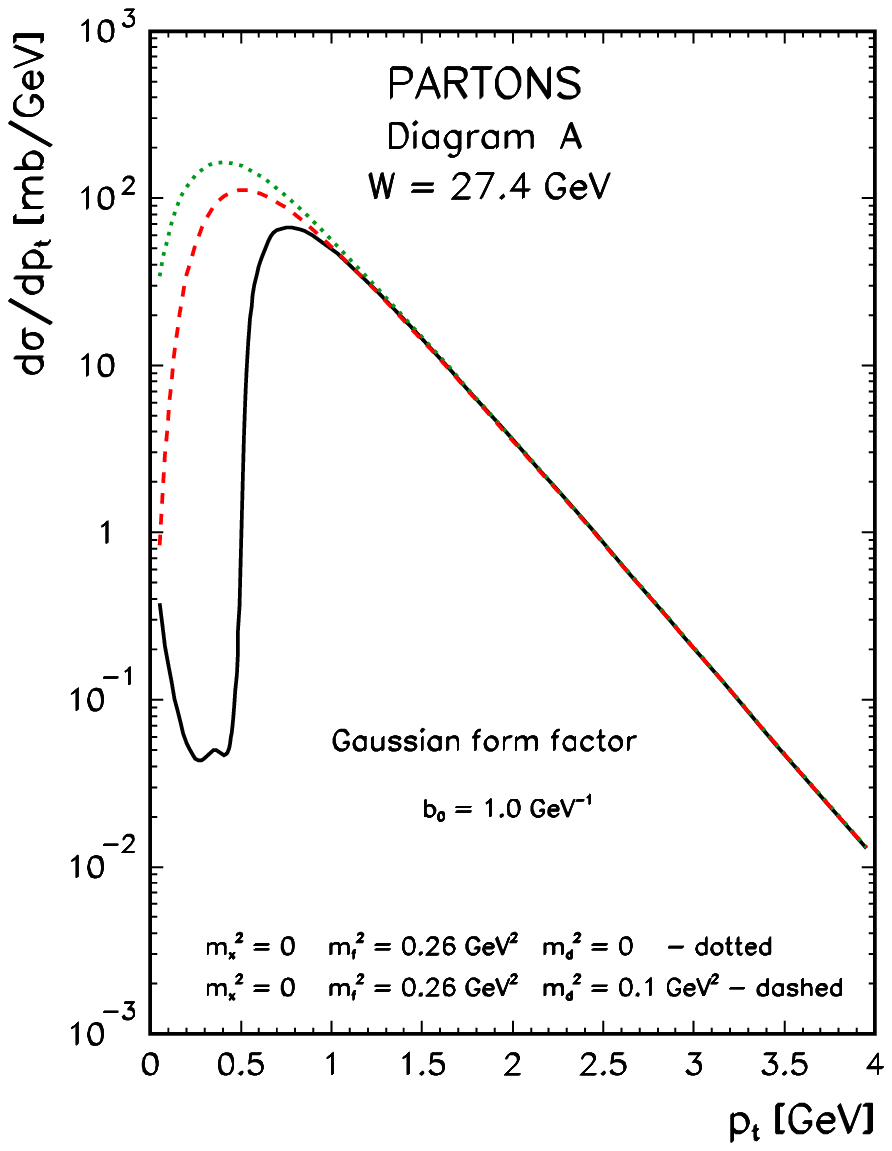}}
  \subfigure[]{\label{partons_b_pt}
    \includegraphics[width=7.0cm]{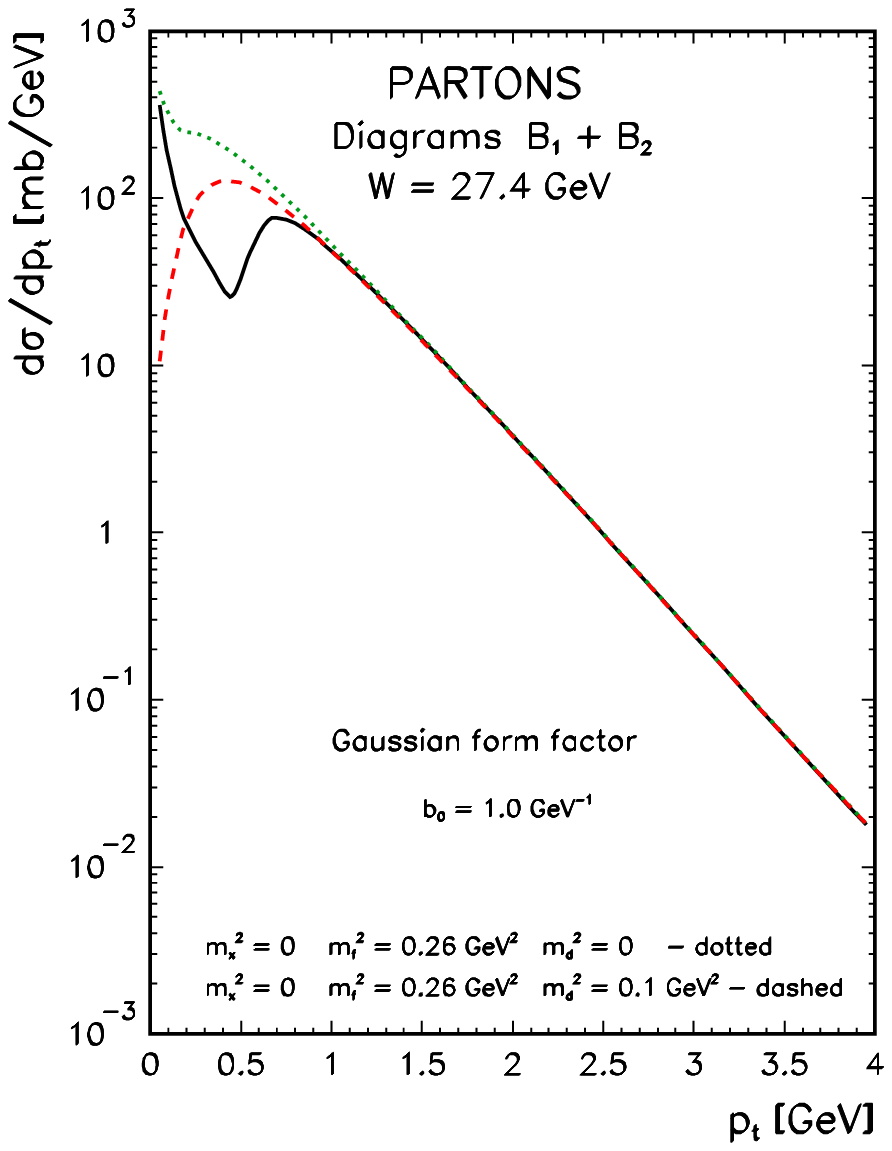}}
\caption{\it
Transverse momentum distribution of partons from diagram $A$ (left panel),
and $B_1 + B_2$ (right panel) for W = 27.4 GeV and -1 $< x_F <$ 1.
In this calculation $b_0$ = 1 GeV$^{-1}$.
The solid line is the same as in the previous figure, i.e. freezing
prescription for the factorization scale was used. For comparison
the dotted line includes the shift prescription for the factorization
scale. Finally the dashed line includes shift of factorization
scale and modification of denominator as described in the text.
\label{fig:partons_pt_extra}
}
\end{figure}

The rise of the cross section above $p_t \approx$ 0.5 GeV
is due to a rapid increase of gluonic radiation above $p_t = \mu_0$,
where $\mu_0^2$ is the minimal factorization scale for the GRV parton
distributions. On the other hand, the rise of the cross section
towards $p_t < \mu_0$ is an artifact
of freezing factorization scale below $\mu_0^2$ in the conjuction
of the singular behaviour of the denominator in
the parton cross section formulae.
To elucidate this issue somewhat better in
Fig.\ref{fig:partons_pt_extra} we present also the results with
the second prescription for the factorization scale with
the minimal shift of the factorization scale
(dotted line) and the result with an extra substitution
of $\frac{1}{p_t^2}$ by $\frac{1}{p_t^2 + m_d^2}$ (dashed line),
where $m_d^2$ is a new nonperturbative parameter of our model
specified in the figure. The second low-scale prescription
lead to a smooth behavior of the partonic cross section
at low transverse momenta.
The consequences and the sensitivity to the extra prescription
on the pion production at low transverse momenta of hadrons will
be discussed in the next section.

\subsection{Pion distributions}

The passage from parton distributions to hadron distributions
is the next step of our analysis. In the present approach we shall use
the approach based on phenomenological fragmentation functions
known from other processes.
In the literature it is mostly $e^+ e^- \to$ hadrons reactions which
are used for extraction of phenomenological fragmentation functions.

There are a few sets of fragmentation functions in the literature.
In the present calculation we shall use the fragmentation function
of Kretzer \cite{Kretzer2000}. There are two advantages of this
particular set of fragmentation functions over other known
from the literature. Firstly, the Kretzer
fragmentation functions are available and reasonable even for very
low scales $p_t \sim$ 1 GeV, or even less, i.e. in the region of
our interest. Secondly, more attention in their construction
was paid to the flavour structure than in any other approach in
the literature. A good control of the flavour structure is essential
when discussing and comparing contributions of diagram
$A$, $B_1$ and $B_2$ to the pion momentum distributions.

\begin{figure}[htb] 
  \subfigure[]{\label{pip_sps_pt}
    \includegraphics[width=7.0cm]{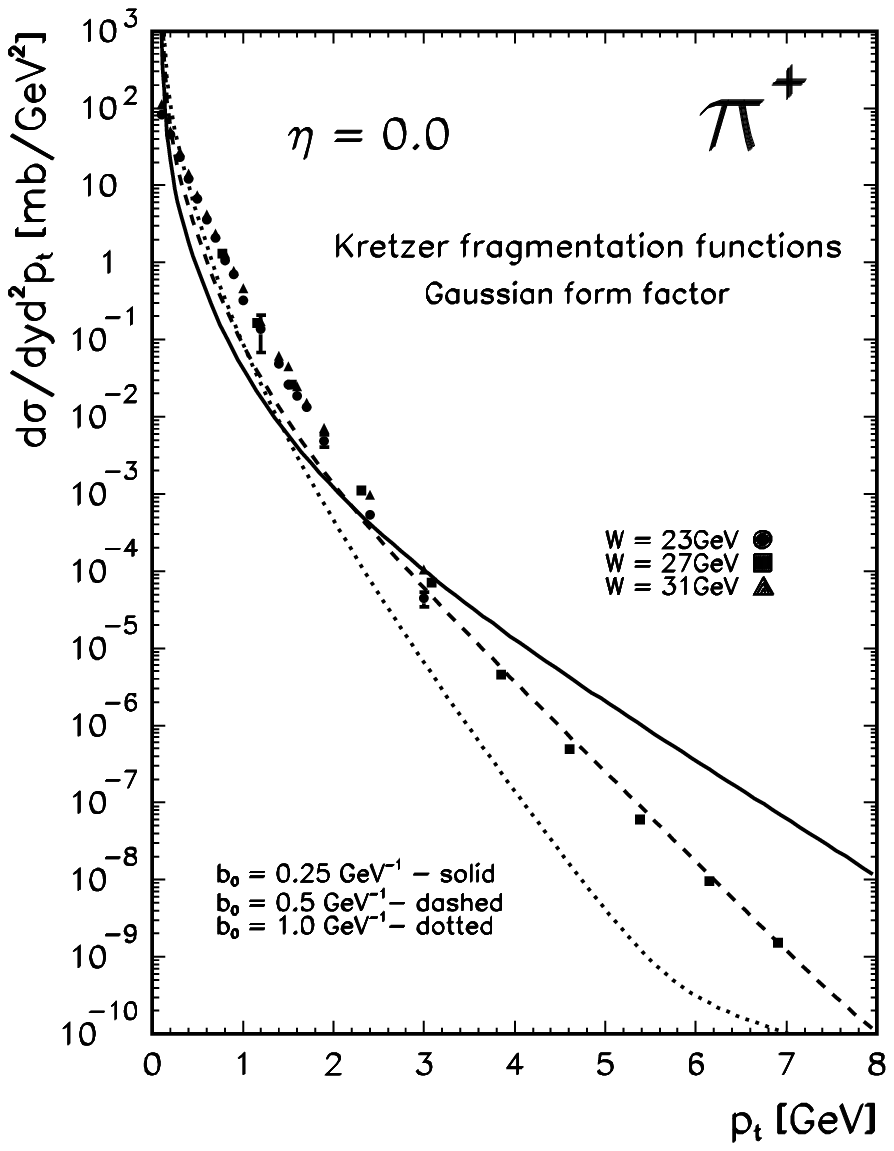}}
  \subfigure[]{\label{pip_sps_pt}
    \includegraphics[width=7.0cm]{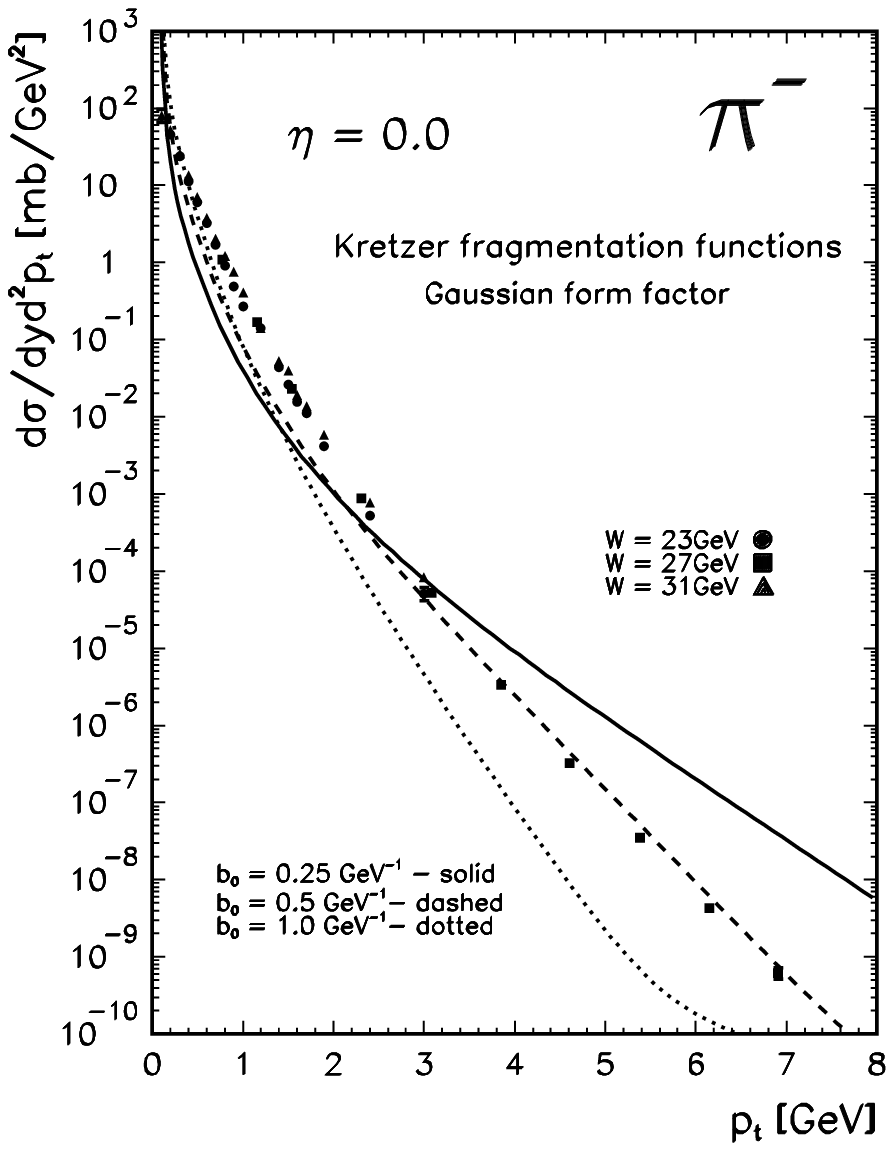}}
\caption{\it
Invariant cross section as a function of transverse momentum
of $\pi^+$ (left panel)
and $\pi^-$ (right panel) for $\eta$ = 0, W = 27.4 GeV and
different values of parameter in the Gaussian form factor.
Experimental data for W  = 23, 31 GeV from \cite{Alper} and
for W = 27.4 GeV \cite{Antreasyan} are shown for comparison.
\label{fig:pions_pt}
}
\end{figure}

In Fig.\ref{fig:pions_pt} we compare our model invariant cross
sections for $p p \to \pi^+$ (left panel) and $p p \to \pi^-$
(right panel) as a function of pion transverse momentum
at W = 27.4 GeV for different values of the parameter $b_0$
of our Gaussian nonperturbative form factor. In principle, our
result should not exceed experimental data especially in
the perturbative regime of $p_t >$ 2 GeV where the perturbative
$2 \to 2$ parton subprocesses are crucial.
This limits the value of the nonperturbative
form factor to $b_0 >$ 0.5 GeV$^{-1}$.  

\begin{figure}[htb] 
  \subfigure[]{\label{pip_sps_small_pt}
    \includegraphics[width=7.0cm]{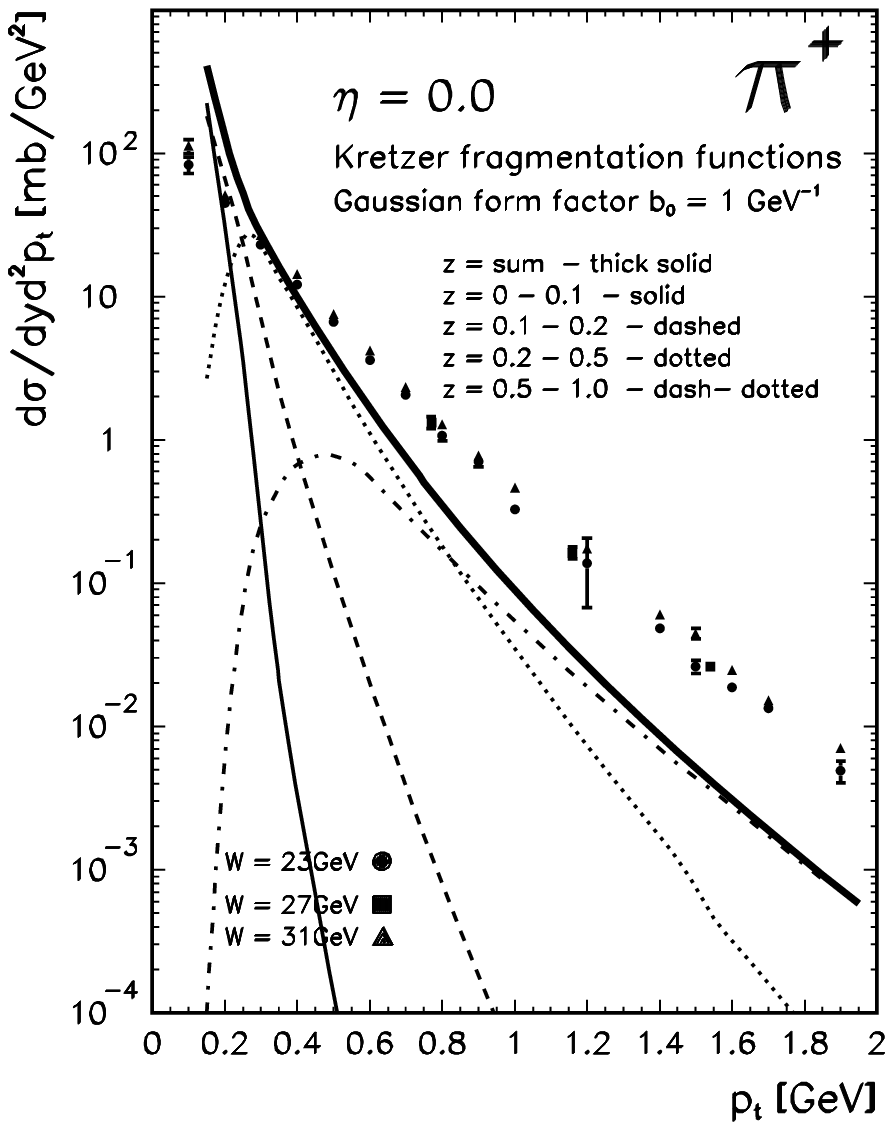}}
  \subfigure[]{\label{pip_sps_small_pt}
    \includegraphics[width=7.0cm]{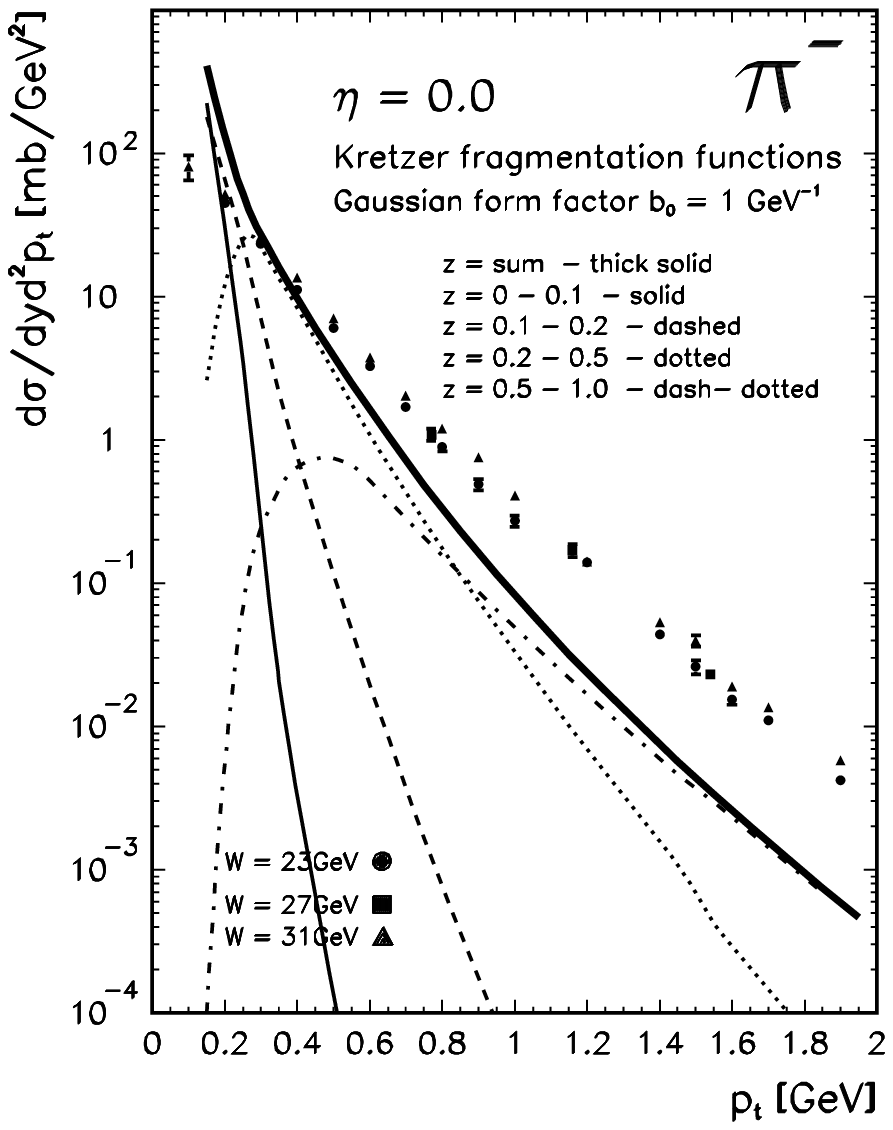}}
\caption{\it
Invariant cross section as a function of
transverse momentum of $\pi^+$ (left panel)
and $\pi^-$ (right panel) for $\eta$ = 0, W = 27.4 GeV.
In this calculation $b_0$ = 1 GeV$^{-1}$.
\label{fig:pions_small_pt}
}
\end{figure}

\begin{figure}[htb] 
  \subfigure[]{\label{gluon_ff}
    \includegraphics[width=4.3cm]{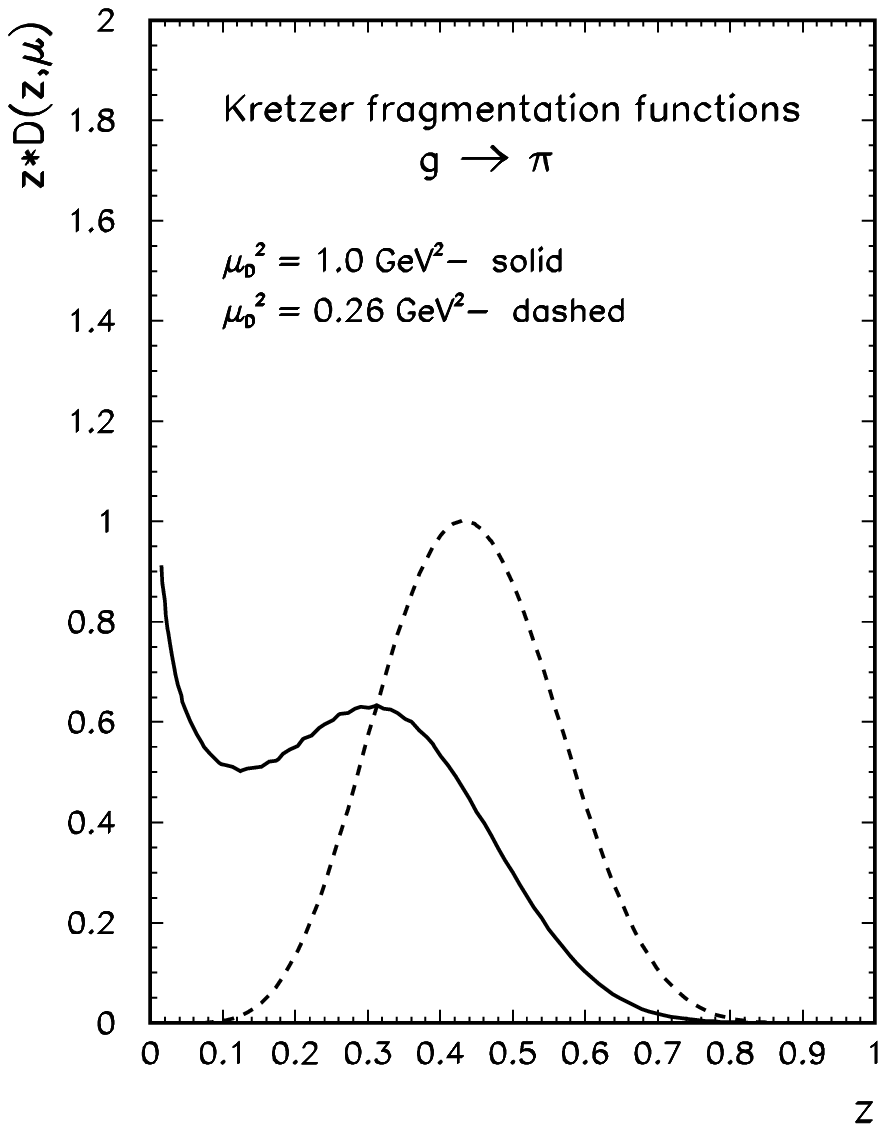}}
  \subfigure[]{\label{favored_ff}
    \includegraphics[width=4.3cm]{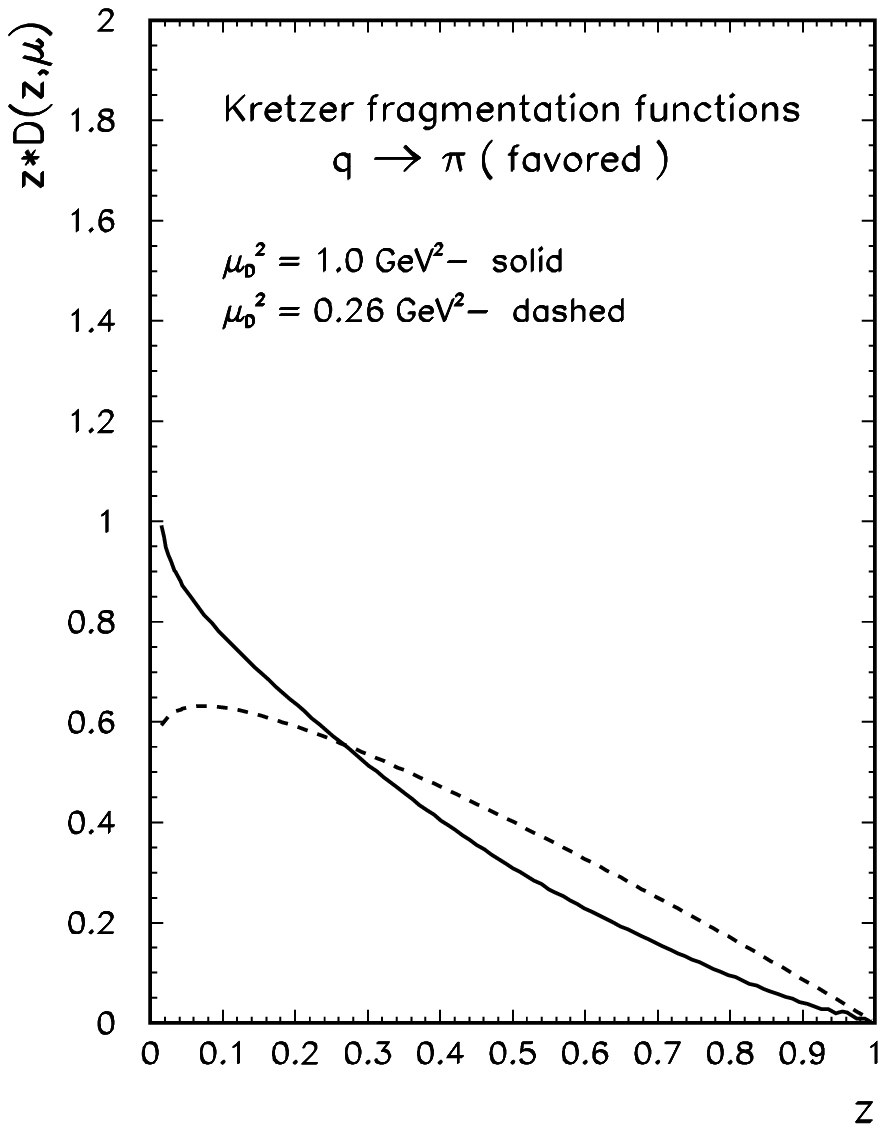}}
  \subfigure[]{\label{unfavored_ff}
    \includegraphics[width=4.3cm]{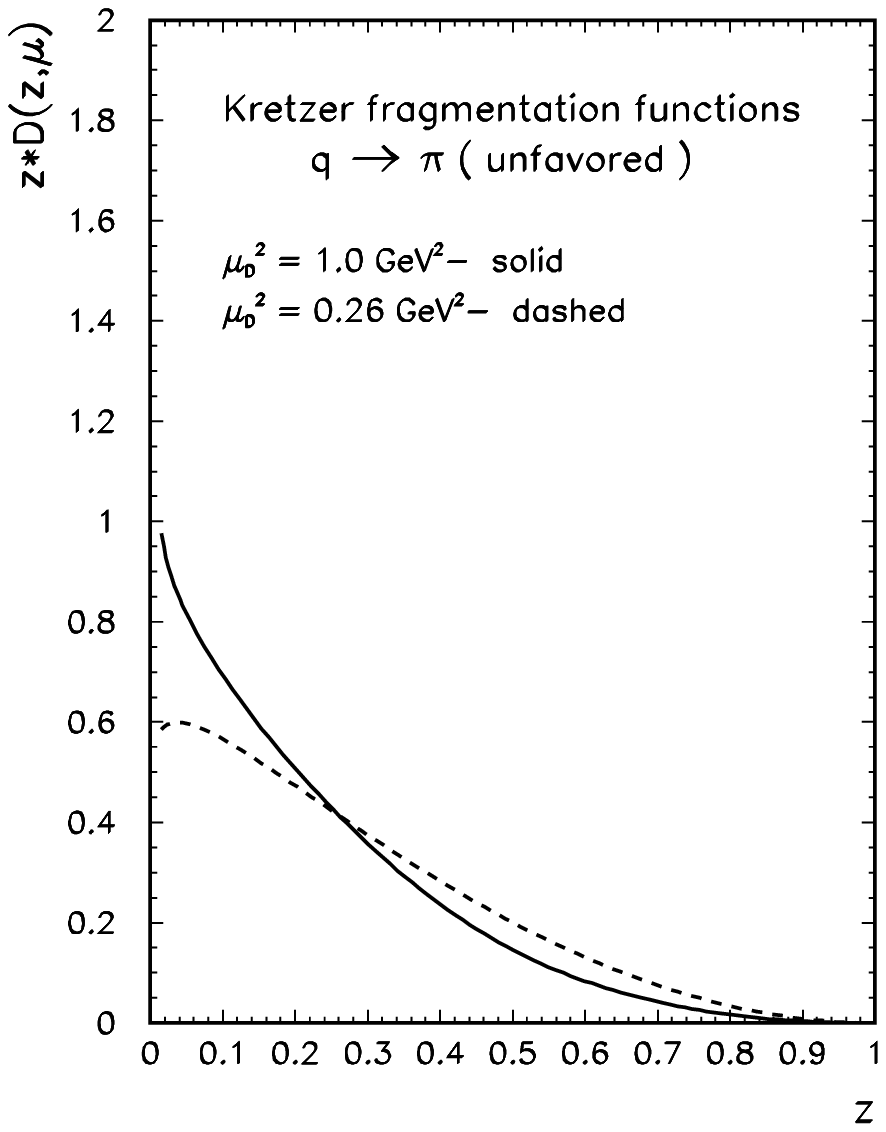}}
\caption{\it
The basic Kretzer fragmentation functions at low factorization
scales.
\label{fig:Kretzer_ff}
}
\end{figure}

\begin{figure}[htb] 
  \subfigure[]{\label{pip_sps_small_pt}
    \includegraphics[width=7.0cm]{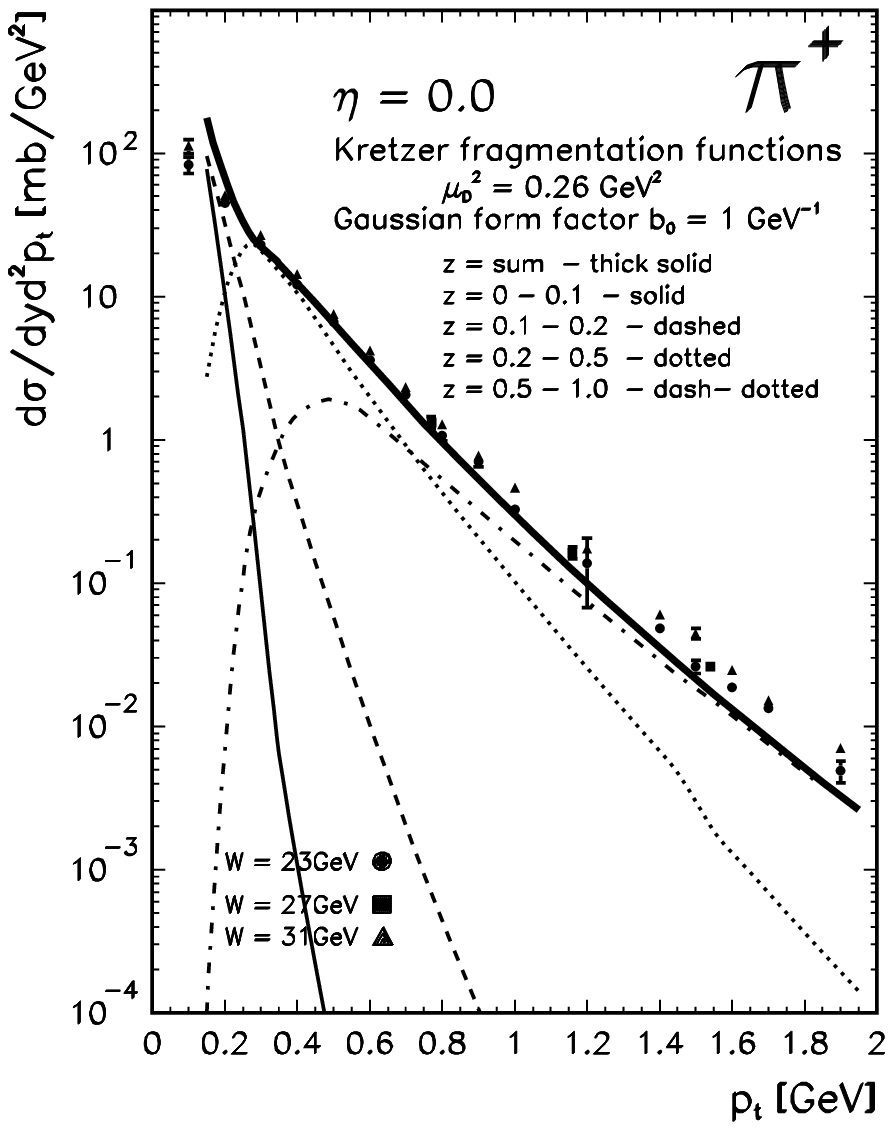}}
  \subfigure[]{\label{pip_sps_small_pt}
    \includegraphics[width=7.0cm]{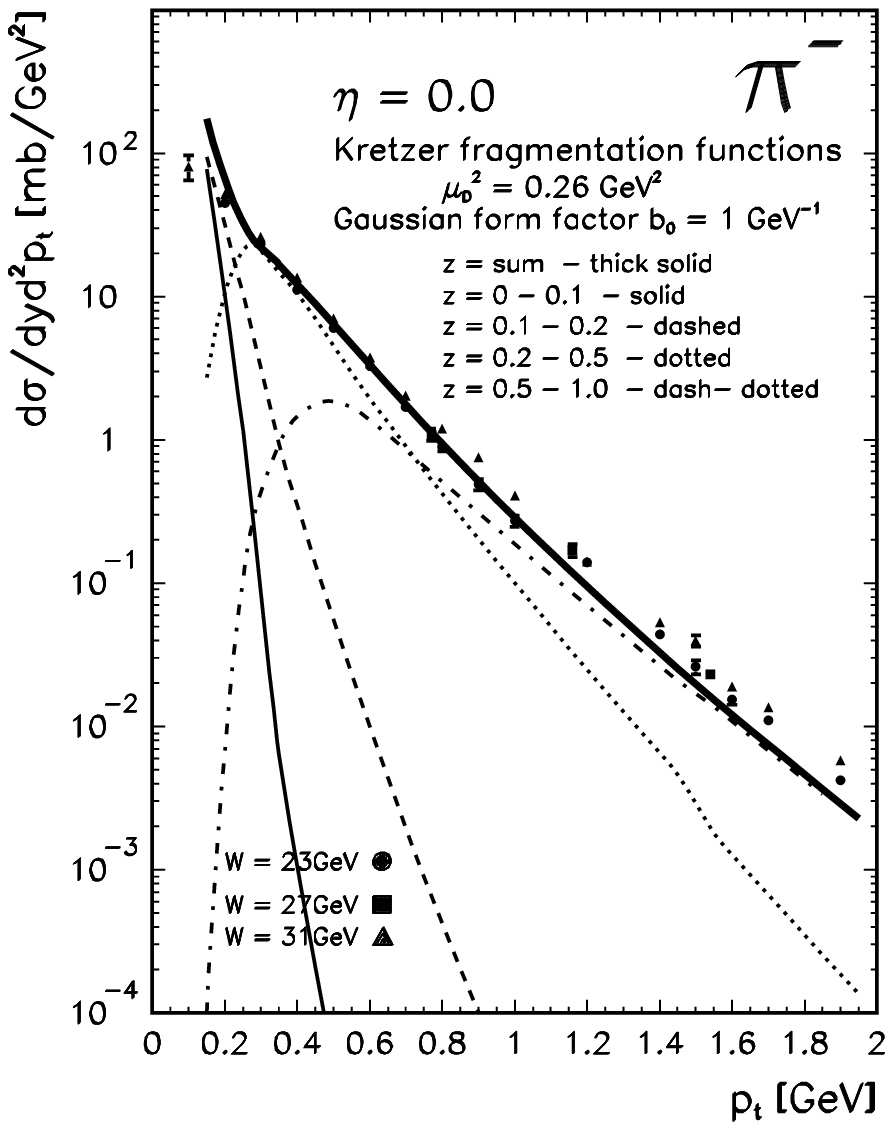}}
\caption{\it
Invariant cross section as a function of
transverse momentum of $\pi^+$ (left panel)
and $\pi^-$ (right panel) for $\eta$ = 0, W = 27.4 GeV obtained
with the Kretzer fragmentation functions at $\mu_0^2$ = 0.26 GeV$^2$.
In this calculation $b_0$ = 1 GeV$^{-1}$.
\label{fig:pions_small_pt_mu20}
}
\end{figure}

As can be seen from the figure the agreement of our model with
experimental data is not perfect. There can be a few reasons for
this fact. In order to understand the problem somewhat better
let us concentrate on a small transverse momentum region
$p_{t,h} <$ 2 GeV. As an example in Fig.\ref{fig:pions_small_pt}
we present our results for $b_0$ = 1 GeV$^{-1}$.
We observe a deficit at $p_t >$ 0.5 GeV and a strong excess
at $p_t <$ 0.3 GeV. It is particularly intriguing which effect
stands behind the huge excess at very small transverse momenta.
In addition, in Fig.\ref{fig:pions_small_pt} we present
contributions of different disjoint and complementary regions of
variable $z$, as indicated in the figure. The figure shows that the rapid
increase below $p_t \sim$ 0.3 GeV is caused exclusively by
small $z <$ 0.2 i.e. must be traced back to the phenomenological
fragmentation functions used. The Kretzer fragmentation functions
as calculated from the publicly available computer code
\cite{Kretzer_private}
are restricted to factorization scales larger than 1
GeV$^2$. Therefore in our calculation we were
forced to freeze the fragmentation functions below $\mu_D = p_t <$ 1 GeV.
The fragmentation functions at this small scale are shown by
the solid line in Fig.\ref{fig:Kretzer_ff} for illustration.
A clear rise of the fragmentation functions at small z can be
observed, which in conjunction with the previous observation
must be responsible for the excess of pions at very small $p_{t,h}$.
The lower limit for the scale of about 1 GeV$^2$ is
a recomendation based on analysis of the $e^+ e^-$ data rather
than a rigorous applicability limit. In fact, the initial evolution
scale in \cite{Kretzer2000} is $\mu_0^2$ = 0.26 GeV$^2$. Let us try to
use the Kretzer fragmentation functions at this even lower scale.
In Fig.\ref{fig:Kretzer_ff} we show (dashed line) the gluon
and quark fragmentation functions. While for $g \to \pi$ fragmentation
the small-z rise in ``z D(z)'' completely dissapears and a minimum at
z = 0 is obtained (valence-like character), for the quark
fragmentation only a saturation of ``z D(z)'' can be observed.
In Fig.\ref{fig:pions_small_pt_mu20} we show the result for pion
production obtained with the fragmentation function at the lowest
factorization scale $\mu_0^2$ = 0.26 GeV$^2$. Clearly the description
of the data is now better. It is somewhat amusing that even
the description above $p_{t,h}$ = 0.5 GeV is now much better.
It becomes obvious that the choice of the factorization scale is
very important in this context. 

The description below $p_{t,h}$ = 0.2 GeV is still rather poor.
This is partially caused by the behaviour of the quark
fragmentation functions at small z (see Fig.\ref{fig:Kretzer_ff}).
Other effects are also possible.  
For example, we have used a simple one-parameter Gaussian form factor.
In principle, such form factor may have more complicated
functional dependence and can depend not only on the impact parameter,
but also on energy and/or $x_1$ and $x_2$. Furthermore
the whole formalism with z-dependent fragmentation functions must
ultimately break at very low transverse momenta, in particular at
small $x_F$ where resonance decays must be treated explicitly.
Having this in view our final result seems promissing
for further improvements in the future.

\begin{figure}[htb] 
\begin{center}
\includegraphics[width=12cm]{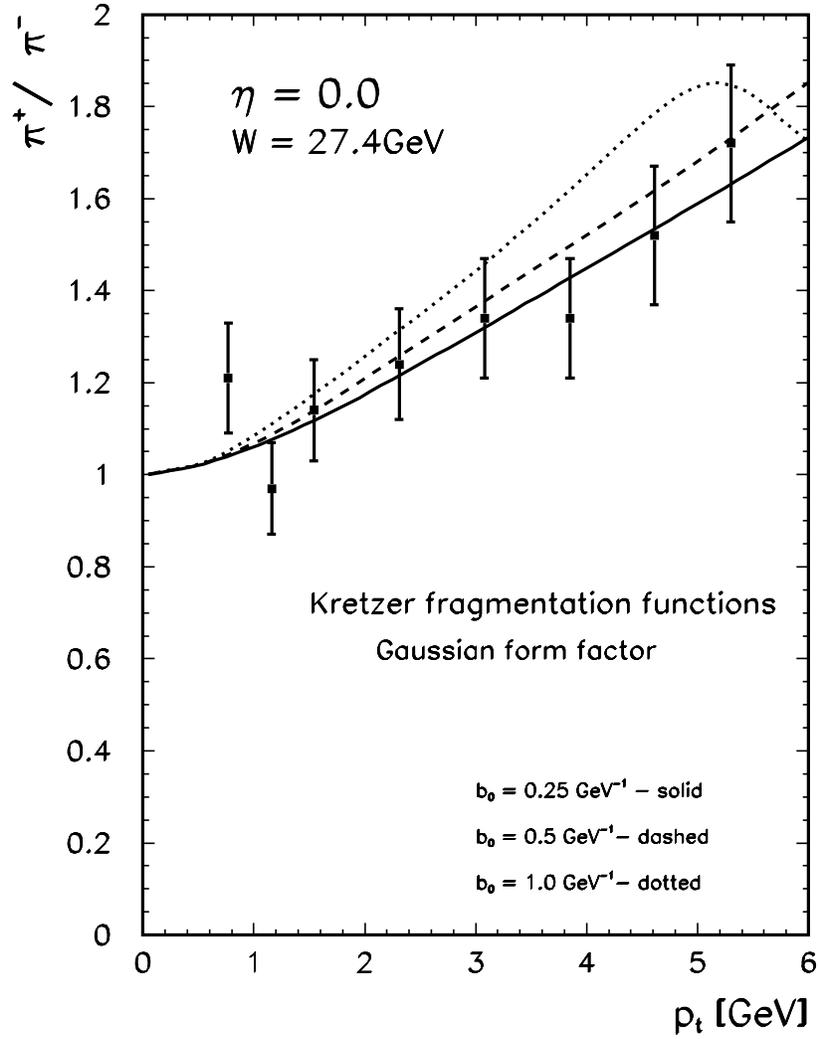}
\caption{\it
Ratio of cross sections from Fig.\ref{fig:pions_pt}
as a function of pion transverse momentum for W = 27.4 GeV.
\label{fig:pim_to_pip_pt}
}
\end{center}
\end{figure}

Inclusion of diagrams $B_1$ and $B_2$ in conjunction
with the flavour dependent fragmentation functions lead to
the $\pi^+ - \pi^-$ asymmetry. In Fig.\ref{fig:pim_to_pip_pt}
we show the asymmetry as the function of pion transverse momentum.
The asymmetry is well described by our model, in contrast to
individual distributions. This seems to suggest the right relative
contributions of diagram $A$, $B_1$ and $B_2$.
The asymmetry depends only weakly on the value of the parameter
$b_0$ of the Gaussian nonperturbative form factor.

\begin{figure}[htb] 
\begin{center}
\includegraphics[width=12cm]{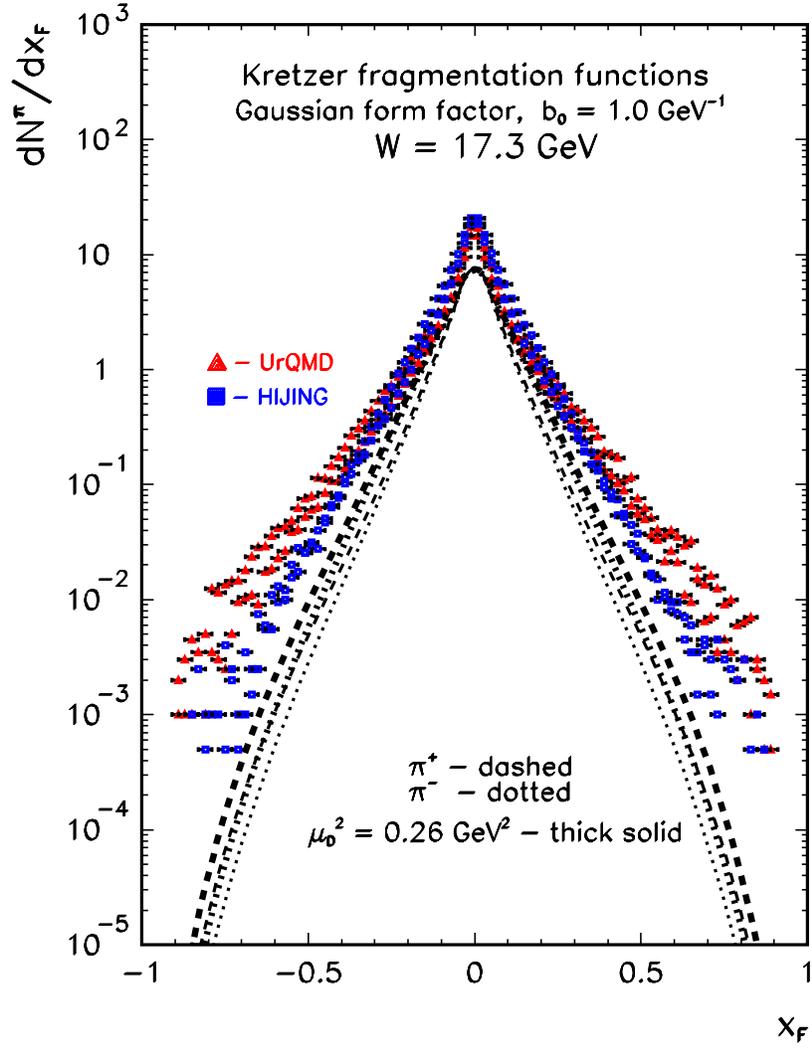}
\caption{\it
Our result against the UrQMD (red triangles)
and HIJING (blue squares) models for SPS energy W = 17.3 GeV,
for positive (dashed) and negative (dotted) pions.
Thin lines correspond to Fig.\ref{fig:pions_small_pt}
and thick lines to Fig.\ref{fig:pions_small_pt_mu20}.
\label{fig:pions_xf}
}
\end{center}
\end{figure}

At SPS energies it is often customary to use $x_F$ rather than rapidity.
The hadron Feynman variable $x_{F,h}$ is related to hadron
pseudorapidity $\eta_h$ as
\begin{equation}
\eta_h = \frac{1}{2} \log \left(
\frac{\sqrt{x_{F,h}^2 \frac{s}{4} + p_{t,h}^2} + x_{F,h} \frac{\sqrt{s}}{2}}
     {\sqrt{x_{F,h}^2 \frac{s}{4} + p_{t,h}^2} - x_{F,h} \frac{\sqrt{s}}{2}}    
\right)
\label{eta_h_x_F_h}
\end{equation}
and the jacobian of transformation expressed by $x_{F,h}$ and
$p_{t,h}$ is
\begin{equation}
\frac{d \eta_h}{d x_{F,h}} 
= \frac{\frac{1}{2} \sqrt{s}}{\sqrt{x_{F,h}^2 \frac{s}{4} +
    p_{t,h}^2}} \; .
\label{eta_h_x_F_h_jacobian}
\end{equation}
In Fig.\ref{fig:pions_xf} we compare the $x_F$ distributions
obtained in our approach for $b_0$ = 1 GeV$^2$ with those
from some popular programs available on the market.
The thin and thick (dashed and dotted) lines correspond
to the different treatment of the scale in the fragmentation functions
as described above (see also the figure caption).
Clearly some mechanisms at large $x_F$ are missing in our approach.
One of such mechanisms is the so-called pion stripping discussed e.g.
in Ref.\cite{pion_stripping}. It was not our goal to include
all of these effects and describe the experimental data.
Our intention here was to identify the dominant mechanisms in
the language of unintegrated parton distributions adequate
at intermediate and low transverse momenta.

\begin{figure}[htb] 
\begin{center}
  \subfigure[]{\label{pip_sps_pt}
    \includegraphics[width=6.5cm]{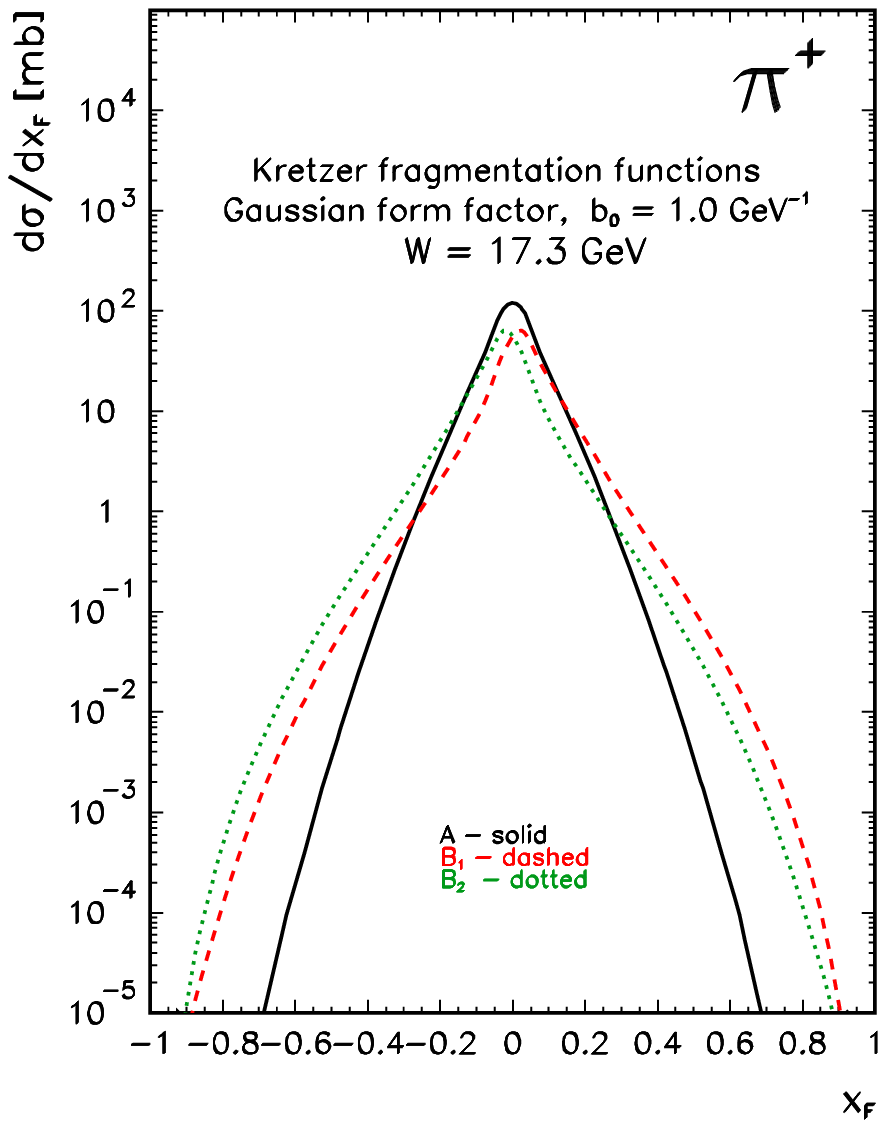}}
  \subfigure[]{\label{pip_sps_pt}
    \includegraphics[width=6.5cm]{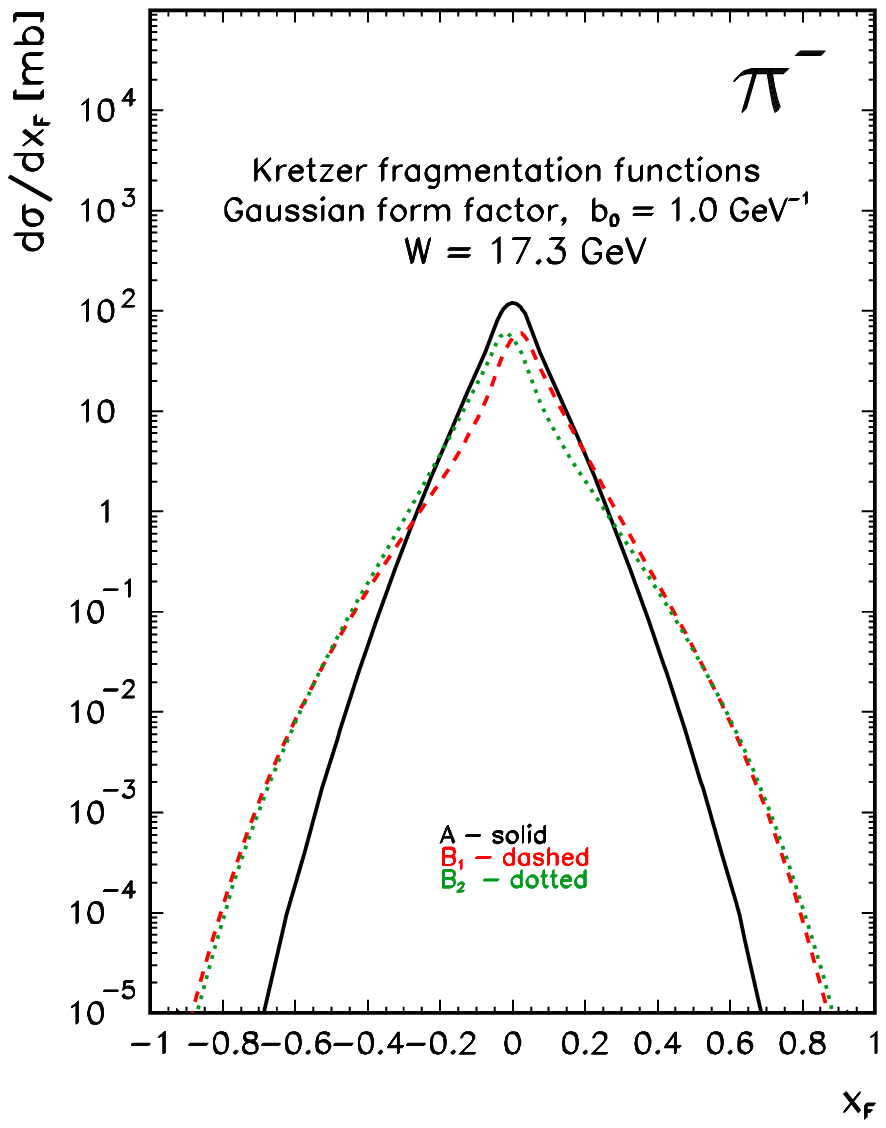}}
\caption{\it
Contribution of diagrams $A$ (thin solid), $B_1$ (dashed)
and $B_2$ (dotted) to charged-pion $x_F$ distributions
for W = 17.3 GeV.
In this calculation $b_0$ = 1.0 GeV$^{-1}$
and 0.2 GeV $< p_{t,h}<$ 2 GeV.
\label{fig:pions_xf_diag}
}
\end{center}
\end{figure}

For completeness in Fig.\ref{fig:pions_xf_diag} we present
a decomposition of the pion $x_F$ distribution into
the components corresponding to diagrams shown in Fig.\ref{fig:diagrams}.
The purely gluonic contribution (diagram $A$) is comparable to
the other two contributions at $x_F \approx$ 0, but at $|x_F| >$ 0.5
the contributions of diagram $B_1$ and $B_2$ are significantly larger.
Similarly as for parton distributions the diagrams $B_1$ and $B_2$
contribute at somewhat larger $|x_F|$ than the diagram $A$.
Some differences for $\pi^+$ and $\pi^-$ distributions are visible
in the figure.
The mechanisms underlying diagrams $B_1$ and $B_2$
are important contributors to understand the $\pi^+ - \pi^-$ asymmetry 
at small and intermediate values of $x_F$.
We think, however, that the dominant mechanism of
$\pi^+$ and $\pi^-$ asymmetry
in very forward ($x_F >$ 0.5) and very backward ($x_F <$ -0.5)
hemispheres is due to the pion stripping mechanism
(for a reference see e.g. \cite{pion_stripping}).

\section{Conclusions}

The approach based on unintegrated gluon distributions was applied
recently to describe particle momentum distributions at RHIC
for nucleus-nucleus \cite{KL01} and proton-proton collisions
\cite{szczurek03}.
We propose new mechanisms, neglected so far in the literature,
which involve also quark/antiquark degrees of freedom
and are based on (anti)quark-gluon and gluon-(anti)quark fusion
processes followed by the subsequent fragmentation.
These missing mechanisms have been estimated in the approach based
on unintegrated parton (gluon, quark, antiquark) distributions
originating from the solution of a set of coupled equations
proposed recently by Kwieci\'nski and coworkers.
The formalism proposed recently by
Kwieci\'nski is very useful to obtain not only gluon unintegrated
distributions but also their counterparts for quarks and antiquarks.

In the present paper we have concentrated on low energies
$W \sim$ 20 GeV, relevant for SPS experiments.
By freezing relevant scales for small parton transverse momenta
we achieve a reasonable description of the data down to pion
transverse momenta $\sim$ 0.5 GeV.
The missing terms lead to an asymmetry in the production
of the $\pi^+$ and $\pi^-$ mesons. While at lower energies of the order
of 20 -- 50 GeV such an asymmetry has been observed experimentally,
it was not yet studied at RHIC. The BRAHMS collaboration at RHIC has
a potential to study the asymmetries. Such asymmetries may be
also useful to pin down the non-gluonic mechanisms as those 
encoded in diagrams $B_1$ and $B_2$.

The non-ideal agreement with the experimental data at low $p_t$ can
be due to approximate treatment of nonperturbative effects embodied in
the form factors, as well as in the treatment of hadronization effects
with the help of scale-dependent one-parameter fragmentation
functions. Both these effects require further detailed studies
which go beyond the scope of the present paper.

{\bf Acknowledgements}
We are indebted to Peter Levai for an interesting discussion
about collinear approach and Stefan Kretzer for a discussion about
parton fragmentation functions.
We are also indebted to Ewa Kozik and Piotr Paw{\l}owski for
performing the calculation with the UrQMD and HIJING code,
respectively. This work was partially supported by the Polish KBN
grant no. 1 P03B 028 28.



\begin{thebibliography}{99}

\bibitem{Owens}
J.F. Owens, Rev. Mod. Phys. {\bf 59} (1987) 465.

\bibitem{Field}
R. D. Field, Application of Perturbative QCD  (Addison-Wesley,
Reading, MA, 1995).

\bibitem{EK97}
K.J. Eskola and K. Kajantie, Z. Phys. {\bf C75} (1997) 515.

\bibitem{EH02}
K.J. Eskola and H. Honkanen, Nucl. Phys. {\bf A713} (2003) 167.

\bibitem{BFLPZ01}
G.G. Barnafoldi, G. Fai, P. Levai, G. Papp, Y. Zhang,
J. Phys. {\bf G27} (2001) 1767.

\bibitem{Wang2000}
X.-N. Wang, Phys. Rev. {\bf C61} (2000) 064910.

\bibitem{Levai_LO}
Y. Zhang, G. Fai, G. Papp, G.G. Barnafoldi and P. Levai,
Phys. Rev. {\bf C65} (2002) 034903.

\bibitem{Levai_NLO}
G. Papp, G.G. Barnafoldi, P. Levai and G. Fai,
hep-ph/0212249.

\bibitem{WW98}
Ch.-Y. Wong and H. Wang, Phys. Rev. {\bf C58} (1998) 376.


\bibitem{KL01}
D. Kharzeev and E. Levin, Phys. Lett. {\bf B523} (2001) 79.

\bibitem{szczurek03}
A. Szczurek, Acta Phys. Polon. {\bf B34} (2003) 3191.

\bibitem{preliminary}
A. Szczurek, Acta Phys. Polon. {\bf B35} (2004) 161.

\bibitem{RHIC}
Proceedings of the Quark Matter 2002 conference, July 2002,
Nantes, France; Nucl. Phys. {\bf A715} (2003). \\
Proceedings of the Quark Matter 2004 conference, January 2004,
Oakland, USA; J. Phys. {\bf G30} (2004).

\bibitem{PHOBOS}
B.B. Back et al. (PHOBOS collaboration), Phys. Rev. Lett. {\bf 87}
(2001) 102303-1.

\bibitem{BRAHMS}
I.G. Bearden et al. (BRAHMS collaboration), Phys. Rev. Lett. {\bf 87}
(2001) 112305.


\bibitem{GLR81}
L.V. Gribov, E.M. Levin and M. G. Ryskin, Phys. Lett. {\bf B100}
(1981) 173.


\bibitem{KMR01}
M.A. Kimber and A.D. Martin and M.G. Ryskin,
Phys. Rev. {\bf D63} (2001) 114027-1.


\bibitem{CCFM_b1}
J. Kwieci\'nski, Acta Phys. Polon. {\bf B33} (2002) 1809.

\bibitem{CCFM_b2}
A. Gawron and J. Kwieci\'nski, Acta Phys. Polon. {\bf B34}
(2003) 133.

\bibitem{GKB03}
A. Gawron, J. Kwieci\'nski and W. Broniowski, Phys. Rev. {\bf D68}
(2003) 054001.

\bibitem{KS04}
J. Kwieci\'nski and A. Szczurek, Nucl. Phys. {\bf B680} (2004) 164.

\bibitem{LS04}
M. {\L}uszczak and A. Szczurek, Phys. Lett. {\bf B594} (2004) 291.


\bibitem{SS97}
D.V. Shirkov and I.L. Solovtsov, Phys. Rev. Lett. {\bf 79} (1997) 1209. 


\bibitem{BKK95}
J. Binnewies, B.A. Kniehl, G. Kramer, Phys. Rev. {\bf D52} (1995) 4947.

\bibitem{Kretzer2000}
S. Kretzer, Phys. Rev. {\bf D62} (2000) 054001.

\bibitem{Kretzer_private}
S. Kretzer, private communication.


\bibitem{Alper}
B. Alper et al. (British-Scandinavian collaboration),
Nucl. Phys. {\bf B100} (1975) 237.

\bibitem{Antreasyan}
D. Antreasyan et al., Phys. Rev. {\bf D19} (1979) 764.


\bibitem{GRV98}
M. Gl\"uck, E. Reya and A. Vogt, Eur. Phys. J. {\bf C5} (1998) 461.

\bibitem{pion_stripping}
N.N. Nikolaev, W. Sch\"afer, A. Szczurek and J. Speth,
Phys. Rev. {\bf D60} (1999) 014004.

\end{thebibliography}
\end{document}